\documentclass{emulateapj}
\usepackage{epsf}
\usepackage{times}
\usepackage{lscape}
\newcommand{\etal}{{et al}\/.}

\def\edit{}

\begin{document}
\slugcomment{Draft of \today} \shorttitle{Long Term Monitoring of the
  Jet of Cen A} \shortauthors{J.L.\,Goodger \etal} \title{Long Term
  Monitoring of the Dynamics and Particle Acceleration of Knots in the
  Jet of Centaurus A} \author{J.L.\ Goodger\altaffilmark{1},
  M.J.\ Hardcastle\altaffilmark{1}, J.H.\ Croston\altaffilmark{3},
  R.P.\ Kraft\altaffilmark{2}, M.\ Birkinshaw\altaffilmark{5,2},
  D.A.\ Evans\altaffilmark{12}, A.\ Jord\'an\altaffilmark{11,2},
  P.E.J.\ Nulsen\altaffilmark{2}, G.R.\ Sivakoff\altaffilmark{8},
  D.M.\ Worrall\altaffilmark{5,2}, N.J.\ Brassington\altaffilmark{2},
  W.R.\ Forman\altaffilmark{2}, M.\ Gilfanov\altaffilmark{9},
  C.\ Jones\altaffilmark{2}, S.S.\ Murray\altaffilmark{2},
  S.\ Raychaudhury\altaffilmark{6}, C.L.\ Sarazin\altaffilmark{8},
  R.\ Voss\altaffilmark{10} and K.A.\ Woodley\altaffilmark{7}}
\altaffiltext{1}{School of Physics, Astronomy \& Mathematics,
  University of Hertfordshire, College Lane, Hatfield AL10 9AB, UK;
  j.l.goodger@herts.ac.uk, m.j.hardcastle@herts.ac.uk}
\altaffiltext{2}{Harvard-Smithsonian Center for Astrophysics, 60
  Garden Street, Cambridge, MA~02138, USA; kraft@head.cfa.harvard.edu,
  nbrassington@cfa.harvard.edu, wforman@cfa.harvard.edu,
  cjones@cfa.harvard.edu, ssm@cfa.harvard.edu,
  pnulsen@cfa.harvard.edu} \altaffiltext{3}{University of Southampton,
  University Road, Southampton SO17 1BJ, UK; J.Croston@soton.ac.uk}
\altaffiltext{4}{European Southern Observatory,
  Karl-Schwarzschild-Str.\ 2 85748 Garching bei M\"{u}nchen, Germany}
\altaffiltext{5}{Department of Physics, University of Bristol, Tyndall
  Avenue, Bristol BS8 ITL, UK; mark.birkinshaw@bristol.ac.uk,
  d.m.worrall@bristol.ac.uk} \altaffiltext{6}{School of Physics and
  Astronomy, University of Birmingham, Edgbaston, Birmingham B15 2TT,
  UK; somak@star.sr.bham.ac.uk} \altaffiltext{7}{Department of Physics
  and Astronomy, McMaster University, Hamilton, ON L8S 4M1, Canada;
  woodleka@physics.mcmaster.ca} \altaffiltext{8}{Department of
  Astronomy, University of Virginia, P.  O. Box 400325,
  Charlottesville, VA 22904-4325, USA; sarazin@virginia.edu,
  grs8g@virginia.edu} \altaffiltext{9}{Max Planck Institute fur
  Astrophysik Karl-Schwarzschild-Str 1 D-85741 Garching, Germany;
  gilfanov@MPA-Garching.mpg.de} \altaffiltext{10}{Excellence Cluster
  Universe Technische Universitt Munchen Boltzmannstr 2 D-85748
  Garching, Germany; rvoss@mpe.mpg.de} \altaffiltext{11}{Departamento
  de Astronom\'{\i}a y Astrof\'{\i}sica, Pontificia Universidad
  Cat\'olica de Chile, 7820436 Macul, Santiago, Chile;
  ajordan@astro.puc.cl} \altaffiltext{12}{MIT Kavli Institute for
  Astrophysics and Space Research, 77 Massachusetts Avenue, Cambridge,
  MA 02139, USA; devans@space.mit.edu}

\begin{abstract}

We present new and archival multi-frequency radio and X-ray data for
Centaurus A obtained over almost 20 years at the VLA and with {\it
  Chandra}, with which we measure the X-ray and radio spectral
indices of jet knots, flux density variations in the jet knots,
polarization variations, and proper motions.  We compare the
observed properties with current knot formation models and particle
acceleration mechanisms.  We rule out impulsive particle acceleration
as a formation mechanism for all of the knots as we detect the same
population of knots in all of the observations and we find no evidence
of extreme variability in the X-ray knots.  We find the most likely
mechanism for all the stationary knots is a collision resulting in a
local shock followed by a steady state of prolonged, stable particle
acceleration and X-ray synchrotron emission.  In this scenario, the
X-ray-only knots have radio counterparts that are too faint to be
detected, while the radio-only knots are due to weak shocks where no
particles are accelerated to X-ray emitting energies.  Although the
base knots are prime candidates for reconfinement shocks, the presence
of a moving knot in this vicinity and the fact that there are two base
knots are hard to explain in this model.  We detect apparent motion in
three knots; however, their velocities and locations provide no
conclusive evidence for or against a faster moving `spine' within the
jet.  The radio-only knots, both stationary and moving, may be due to
compression of the fluid.

\end{abstract}


\section{Introduction}
\label{introjlg}

It is generally agreed that the observed emission from Fanaroff-Riley
class I \citep[FR\,I;][]{fr74jlg} radio jets is due to the synchrotron
process at all wavelengths, with similar jet structure observed from
the radio through the optical into the X-ray
\citep[e.g.][]{hardcastle02jlg,harris02jlg}.  The jets of FR\,I radio
galaxies are thought to decelerate as they move away from the core,
entraining material and expanding into a plume of diffuse matter
\citep[e.g.][]{bicknell84jlg}.  One of the most significant
implications of the synchrotron emission model is reflected in the
characteristic loss timescales: in a stable environment, the X-ray
emitting electrons have lifetimes of the order of tens of years,
tracing regions of current, in situ particle acceleration, while the
radio emitting electrons last for hundreds of thousands of years,
showing the history of particle acceleration in the jet.

In order to investigate these regions of particle acceleration we need
data with sensitivity and resolution sufficient to detect jet
substructure on spatial scales comparable to the synchrotron loss
scales.  This prompts us to look to the two closest bright FR\,I radio
jets: M87 and Centaurus A (NGC\,5128, hereafter Cen A).  Both of these
jets have been detected in multiple frequencies from the radio through
to the X-ray
\citep[e.g.][]{feigelson81jlg,kraft02jlg,hardcastle03jlg,hardcastle06jlg,harris02jlg}.
The proximity of these radio galaxies, 16.7\,Mpc and 3.7\,Mpc
respectively \citep{blakeslee09jlg,mei07jlg,ferrarese07jlg}, make them
unique jet laboratories with spatial scales of 77\,pc and 17\,pc per
arcsec respectively.  The details revealed in the structure of these
jets have been the focus of many recent studies
\citep[e.g.][]{biretta99jlg,hardcastle03jlg,kataoka06jlg,cheung07jlg}.
Within the smooth surface brightness observed in both of these jets
are clumps of bright material -- the knots -- embedded in diffuse
material, all emitting via synchrotron emission.  The precise
mechanisms causing the particle acceleration responsible for the
diffuse structure and the knots are still unknown.

The most surprising result of recent observations of these two
  systems was the radio-to-X-ray synchrotron flare of HST-1 in M87.
In 2002, the X-ray flux of HST-1 increased by a factor of 2 in only
116 days \citep{harris03jlg}, implying a change within an emitting
volume with a characteristic size less than 0.1\,pc for a stationary
source (much less than the size of HST-1, $\sim 3$\,pc).  The X-ray
brightness then faded in the following months only to flare again,
peaking in 2005.  At its brightest, this flare was higher than its
2001 level by a factor of $\sim50$.  The UV and radio light-curves
were found to vary in step with the X-ray up to this peak
\citep{perlman03jlg,harris06jlg}, but the subsequent decrease appeared
to drop off faster in the X-ray than in either the optical or the UV,
which drop off in step \citep{harris09jlg}.  In additional to this
spectral variability, it was established with the {\it Hubble Space
  Telescope (HST)} and the NRAO Very Long Baseline Array (VLBA) that
some of the knots in M87 move superluminally, including subregions of
the HST-1 knot \citep{biretta99jlg,cheung07jlg}.  Together, this
suggests that we are observing synchrotron losses in addition to
either beaming or compression/rarefaction of the fluid.

Cen A is a factor of 4.5 closer than M87 so we can resolve more
details in the complicated fine structure of the jet.  The originally
identified features, named A-G by \citet{feigelson81jlg}, have since
been resolved into at least 40 individual knots
\citep{kraft02jlg,hardcastle03jlg} with additional emission from
diffuse material.  Some of the diffuse emission has been described as
downstream `tails' of emission from the knots \citep{hardcastle03jlg}
or as evidence for limb-brightening of the jet \citep{kraft00jlg}.  In
2003, Hardcastle et al. presented 8.4\,GHz radio observations from the
NRAO Very Large Array (VLA) of Cen A.  That work used archival data
from 1991 and new observations from 2002 to study the jet knots and
investigated the offsets and relationships between the radio knots and
their X-ray counterparts and vice versa.  They found that only some of
the radio knots appeared to have X-ray counterparts, leaving many as
`radio-only' knots and `X-ray-only' knots.  They also considered the
temporal changes in the radio knots, specifically their proper
motions, finding that some of the radio knots were moving.  These
moving knots had comparatively little X-ray emission suggesting that
high-energy particle acceleration is less efficient in these regions
than in the jet as a whole.

Some of the current models explaining the presence of knots within the
generally smooth diffuse material of the jet include compressions in
the fluid flow, collisions with obstacles in the galaxy causing local
shocks, reconfinement of the jet or some other jet-wide process, and
magnetic reconnection.  \citet{hardcastle03jlg} ruled out simple
compression of the fluid as a mechanism for producing X-ray bright,
radio faint compact knots in favor of in situ particle acceleration
associated with local shocks; however, compression could still play a
part in the other knots.  They concluded that the most likely model to
describe the majority of these knots is an interaction between the jet
fluid and an obstacle such as a molecular cloud or a high mass-loss
star.  By exploring the temporal behavior of the X-ray and radio
emission, we can understand the evolution of the knots and constrain
the various models of particle acceleration used to describe the jet
features.

In this work we use Chandra and VLA data spread over almost 20 years
to measure the X-ray and radio spectral indices knots, flux density
variations, polarization variations, and the proper motions of the jet
knots in Cen A. Our aims are to detect variability in the radio and
X-ray properties of the knots, either extreme variability similar to
that of HST-1 in M87 or more subtle changes, and to compare these
properties. which will allow us to constrain the knot formation
processes at work in the jet of Cen A.  The details of our radio and
X-ray data reduction are discussed in Section~\ref{sec:data}.  In
Section~\ref{sec:res} we discuss the details of our analysis methods
and the global results for the knot population, highlighting
particularly interesting features.  In Section~\ref{sec:dis}, we
compare these knot properties with the predictions of various models
for the formation of knots, their particle acceleration and the jet
structure.  Finally we outline the most likely processes for forming
knots in Cen A in Section~\ref{sec:con}.

\section{Data}
\label{sec:data}

\begin{figure*}
\includegraphics[width=0.94\textwidth]{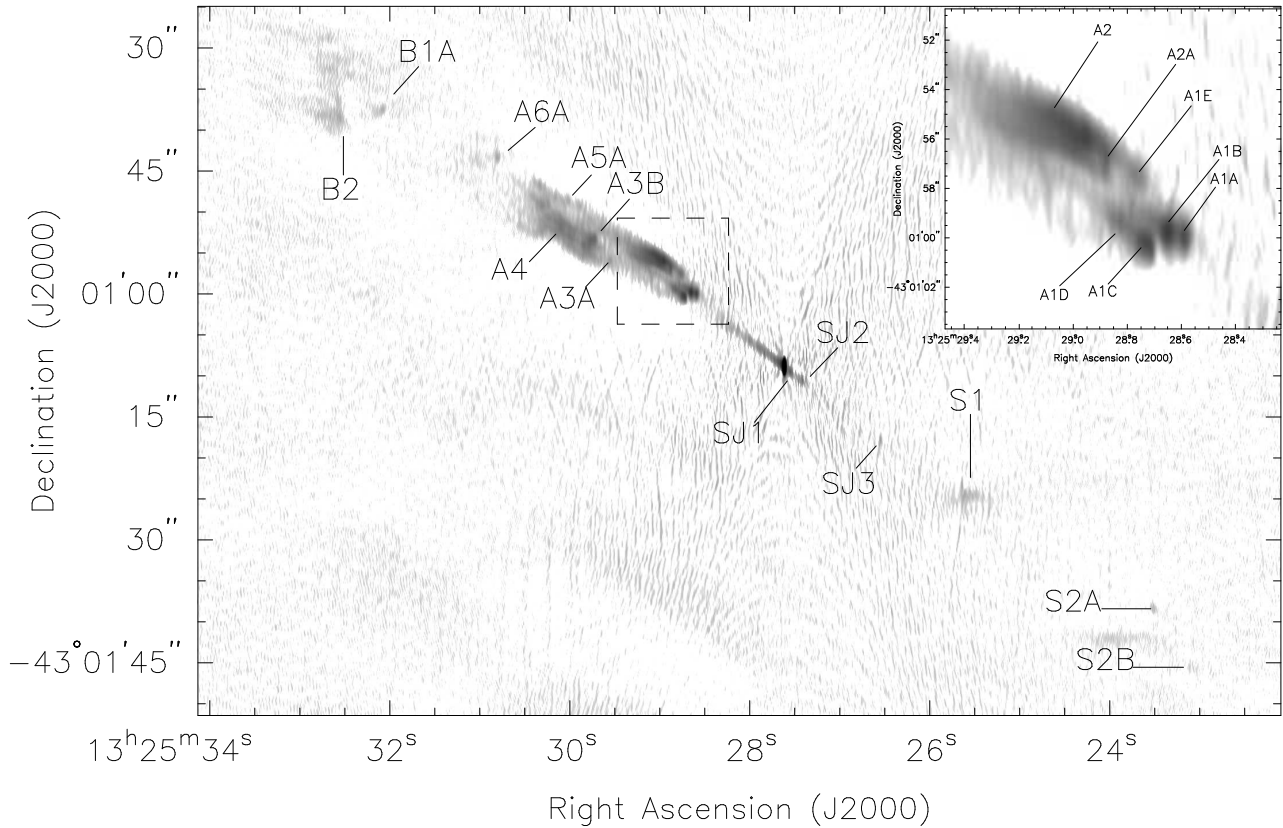}
\caption{Stacked 8.4\,GHz A-configuration radio image of the inner jet in Cen A showing the jet and counterjet with the 19 radio knots labelled as well as the A2 diffuse region.  The dashed box indicates the location of the insert panel (top left), which shows the A1 and A2 groups of knots.  In both the main image and the insert, black corresponds to 0.1\,Jy beam$^{-1}$; however, white corresponds to 0.07\,mJy beam$^{-1}$ and 0.2\,mJy beam$^{-1}$ respectively.  The beam is $0.2\times0.8$\,arcsec.}
\label{fig:radiojlg}
\end{figure*}

In this work we use both new and archival VLA radio data at 4.8\,GHz,
8.4\,GHz and 22\,GHz observed over almost 20 years.  Cen A has been
observed at 4.8\,GHz with the VLA since 1983 and at 8.4\,GHz since
1991, including 6 monitoring observations taken by us since 2002 at
roughly 18 month intervals.  We also present new 22\,GHz data taken in
2007 as part of a multi-frequency program (AG0754) where
quasi-simultaneous observations of the jet knots were taken in these
three radio frequencies timed to coincide with {\it Chandra} X-ray
observations.  The details of these radio data are shown in
Table~\ref{tab:obsjlg}.  All these data were observed with two IFs
with beamwidths of 50\,MHz.  The radio frequencies used in this paper
and in Table~\ref{tab:obsjlg} are the average values for the two IFs.
All the radio data were reduced in \textsc{aips} using the standard
method.  The data were phase-, flux- and polarization-calibrated
before being split into a single source file.  As our flux calibrator,
3C286, is resolved we followed the recommended method of using a model
during the flux calibration.  For an initial approximation, the Cen A
data were calibrated with a point source model before being self
calibrated in phase to the point where no further improvement was
noticed.  The data were then amplitude and phase self-calibrated and
baseline calibrated using the same method as
\citet{hardcastle03jlg}. The radio data were translated to the
published coordinates for the core of Cen A \citep{ma98jlg} in the uv
plane using {\it uvfix} and {\it puthead}.  The radio jet with all the
radio knots labelled is shown in Figure~\ref{fig:radiojlg}; this image
is the combination of all the 8.4\,GHz radio images, individually
scaled by the weighted mean of the rms background.

\begin{figure*}
\includegraphics[width=0.94\textwidth]{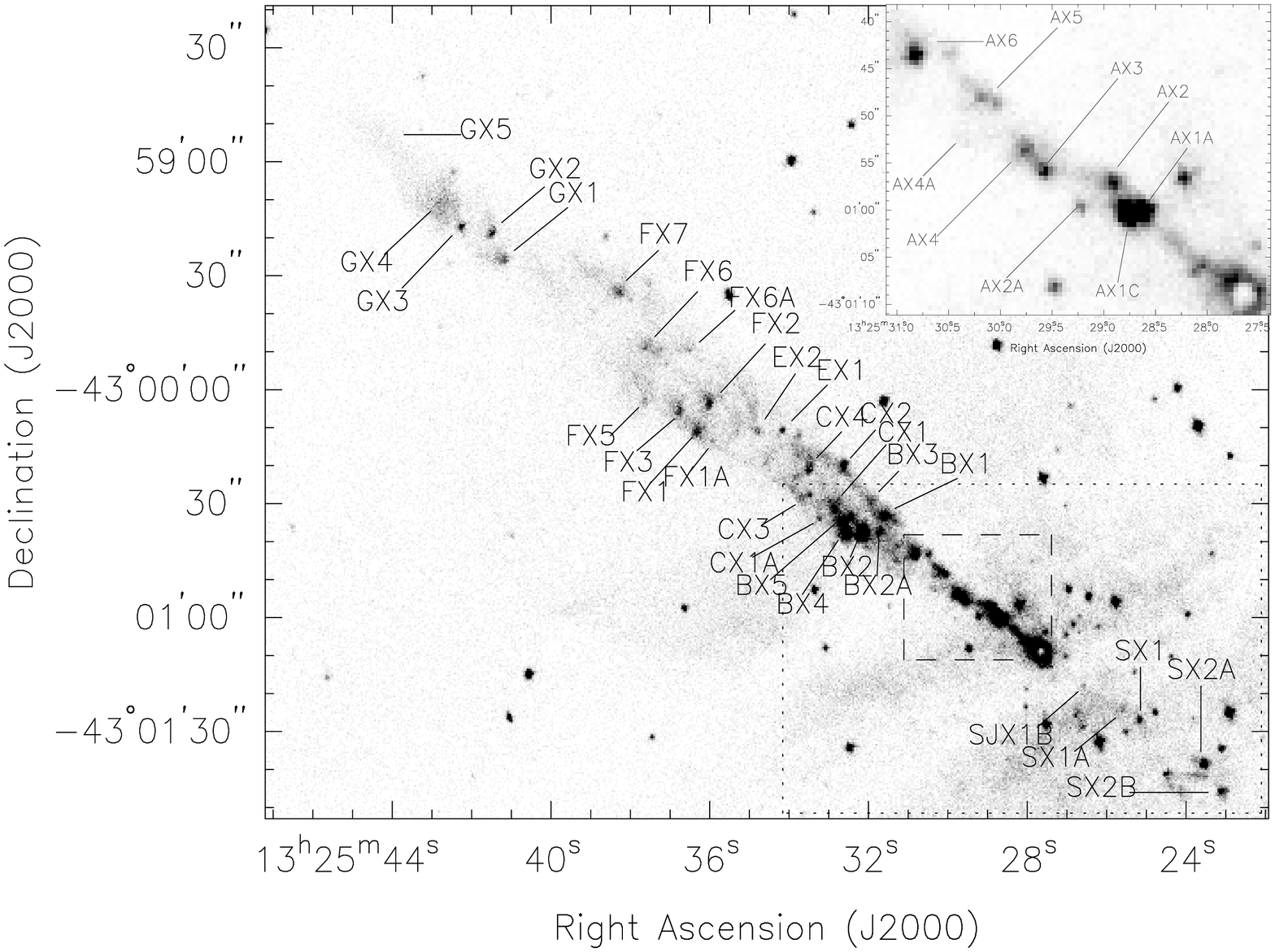}
\caption{X-ray image of the jet in Cen A with all 40 detected X-ray knots labelled.  The X-ray image is in the energy range 0.4 -- 2.5\,keV and shows only the portion of the data which includes the jet and counterjet. The dotted line indicates the extent of the radio jet shown in Figure~\ref{fig:radiojlg} and the dashed line indicates the position of the insert (top left), which shows the A group of X-ray knots.  In the larger image, black corresponds to 28 counts/pixel (0.4 -- 2.5\,keV) and in the insert, black corresponds to 150 counts/pixel (0.4 -- 2.5\,keV). In both images, white corresponds to 0 counts/pixel.  In both images, the pixel size is 0.07 arcsec/pixel. }
\label{fig:xrayjlg}
\end{figure*}

Cen A has also been observed in the X-ray with {\it Chandra} 10 times
since 1999 (observations summarized in Table~\ref{tab:obsxray}).
These X-ray data span 8 years and were taken in such a way that the
jet is unaffected by the chip gaps or the read-out streak of the core.
With the high resolution provided by {\it Chandra}, these well sampled
data give us a unique opportunity to study the temporal properties of
the jet and its knots.  The most recent data, taken in 2007, were part
of a {\it Chandra} Very Large Program (VLP: P.I. R.P. Kraft), consisting of
$6 \times 100$\,ks observations, giving us a combined livetime of
719\,ks when merged with the earlier data. A summary of the reduction
processes is given in \citet{sivakoff08jlg} and
\citet{hardcastle07jlg}.  These VLP data have been used thus far to
study the X-ray binaries in Cen A
\citep{jordan07jlg,sivakoff08jlg,voss09jlg}, the properties of its hot
gas \citep{kraft08jlg,croston09jlg} and of its jet
\citep{hardcastle07jlg,worrall08jlg}.  The merged X-ray data set, in
the energy range 0.4 -- 2.5\,keV, is shown in Figure~\ref{fig:xrayjlg}
with the X-ray knots labeled.

X-ray spectra were extracted for each of the jet knots (see
Section~\ref{sec:res}) and the diffuse regions in the jet, using the
\textsc{ciao} task {\it specextract}, which also calculated the
response files, while \emph{psextract} was used to extract the spectra
of all the point sources in the field for a comparative sample
(Section~\ref{sec:flux-var}).  As some of the jet knots are clearly
extended in the X-ray data (and we wanted to compare with other more
extended jet features), we used {\it specextract} (appropriate for
extended sources) for all of the jet knots for consistency; however,
we used {\it psextract} for the comparison sample, as it consisted
entirely of unresolved point sources.  As the jet knots are compact
sources embedded in the diffuse jet and lobe material, local,
on-source annular regions were used for the background subtraction.
The spectra were then binned to 20 counts per channel after background
subtraction, ignoring the first 28 channels that correspond to
energies below 0.4\,keV.  The X-ray fitting was carried out using
\textsc{xspec} 11.3 in the energy range 0.4 -- 7.0\,keV where the {\it
  Chandra} response is well calibrated.  In this analysis, {\it
  Chandra} data processing was done using CIAO version 3.4 and CALDB
version 3.3.0.1.  We define the spectral index such that flux density,
$S \propto \nu^{-\alpha}$ and the photon index as $\Gamma = 1 +
\alpha$.  The errors in this work are $1\sigma$ errors unless
otherwise stated.

\begin{table*}
  \centering
    \caption{Radio Observation Details}
    \begin{tabular}{clcccccccr}
      \hline
      Frequency & VLA & Date & Program ID &Dynamic Range & Res. & rms Noise\\
       (GHz)  & Config. & & & (arcsec) & (mJy/beam)\\
      \hline
        4.8164 & A & 28-Oct 1983 & AB0257 & 18300:1 &  1.23 $\times$ 0.33 & 0.042\\[3pt]
	4.8164 & A & 16-Mar 1986 & AF0113 & 6570:1 &  1.41 $\times$ 0.36 & 0.264\\[3pt]
	4.8851 & A & 18-Dec 1992 & AK0316 & 27300:1 &  1.63 $\times$ 0.38 & 0.025\\[3pt]
	4.8851 & A & 07-Jun 2007 & AG0754 & 14800:1 &  1.58 $\times$ 0.38 & 0.349\\[3pt]
      \hline
        8.4399 & A & 02-Jan 1991 & AB0587 & 91800:1 & 0.78 $\times$ 0.20 & 0.071\\[3pt]
        8.4601 & A & 03-Mar 2002 & AH0764 & 95700:1 & 0.76 $\times$ 0.20 & 0.067\\[3pt]
        8.4601 & A+PT & 02-Jun 2003 & AH0813 & 32300:1 & 0.79 $\times$ 0.21 & 0.233\\[3pt]
        8.4601 & A+PT & 14-Dec 2004 & AH0855 & 77300:1 & 0.76 $\times$ 0.19 & 0.082\\[3pt]
        8.4601 & A+PT & 18-Feb 2006 & AH0892 & 67500:1 & 0.65 $\times$ 0.15 & 0.103\\[3pt]
        8.4601 & A & 04-Jun 2007 & AG0754 & 92200:1 & 0.82 $\times$ 0.23 & 0.069\\[3pt]
	8.4061 & A & 20-Dec 2008 & AG0798 & 78800:1 & 0.82 $\times$ 0.21 & 0.069\\[3pt]
      \hline
        22.4851 & A & 16-Jun 2007 & AG0754 & 14800:1 & 0.72 $\times$ 0.19 & 0.457\\
	        & B & 21-Dec 2007 & & & & \\[3pt]
      \hline
   \end{tabular}
  \label{tab:obsjlg}
\end{table*}

\begin{table}
  \centering
    \caption{X-ray Observation Details}
    \begin{tabular}{cccc}
      \hline
        Obs.  & Date & Detector & Exposure \\
	(\#) & & & (s) \\
      \hline
        0316 & 05-Dec 1999 & ACIS-I & 26493\\
	0962 & 17-May 2000 & ACIS-I & 36505\\ 
	2978 & 03-Sep 2002 & ACIS-S & 44589\\
	3965 & 14-Sep 2003 & ACIS-S & 49518\\ 
	7797 & 22-Mar 2007 & ACIS-I & 96888\\
	7798 & 27-Mar 2007 & ACIS-I & 90839\\
	7799 & 30-Mar 2007 & ACIS-I & 94783\\
	7800 & 17-Apr 2007 & ACIS-I & 90843\\
	8489 & 08-May 2007 & ACIS-I & 93936\\
	8490 & 30-May 2007 & ACIS-I & 94425\\
      \hline
   \end{tabular}
  \label{tab:obsxray}
\end{table}

\section{Results}
\label{sec:res}

Our radio and X-ray data sets allow us to analyze the temporal
behavior of the knots in Cen A's jet.  We first consider whether the
knots are really jet related or whether some of them are
coincidentally positioned foreground or background objects
(Section~\ref{sec:ps}).  Next, in Section~\ref{sec:em}, we examine the
multi-frequency radio and X-ray data to determine whether
inverse-Compton emission (iC) is significant in the jet knots before
establishing the proper motions of the knots in Section~\ref{sec:pm}.
We then consider the radio, polarization, and X-ray variability of the
knots (Sections~\ref{sec:flux-var} and~\ref{sec:pol}).  We also
measure the radio spectral indices, X-ray spectral indices and
X-ray/radio flux density ratio to determine the broad spectra of the knots
(Sections~\ref{sec:index} and~\ref{sec:counters}).  Combining these
properties provides us with evidence to test models for particle
acceleration in the jet. We also investigate whether any of the knot
properties depend on the position of the knot in the jet, following up
previous work by \citet{hardcastle07jlg} and \citet{worrall08jlg}.

We combined all of our 8.4\,GHz radio data and all the X-ray
observations to make deep, high dynamic range, radio and X-ray maps
shown in Figures~\ref{fig:radiojlg} and~\ref{fig:xrayjlg}.  These maps
allowed us to make a definitive list of all the radio and X-ray knots
in the jet.  The 19 radio knots investigated in this work are mostly
those defined by \citet{hardcastle03jlg} with the addition of two
knots, located downstream of the previously detected A1 knots.  These
knots were present in previous observations but were considered to be
diffuse downstream emission. However, in the more recent observations,
they appear much more compact, so have been designated A1D and A1E and
are investigated in this work.  In the 8.4\,GHz radio maps, where the
resolution is $0.8 \times 0.2$\,arcsec, we find bandwidth smearing
significantly affects knots beyond 140\,arcsec which is beyond the
F-group of X-ray knots so does not affect the radio knots in the
A-configuration maps.  We also note that the radio jet is within the
primary beam of the VLA at all of our observed frequencies so primary
beam attenuation is not corrected for.  Time-averaging smearing is
also not significant at these scales.  We examined maps which extend
to the inner edge of the inner lobes \citep[including the B-array VLA
  data of][]{hardcastle03jlg} and we find no evidence of additional
compact radio knots beyond the B-group radio knots already detected.
Bandwidth smearing does not affect these images until beyond the
jet.   The absence of knots at large distances from the nucleus
  will be discussed further in section~\ref{sec:dis-col}. 

The 40 X-ray knots used in this work are a combination of those
identified by \citet{kraft02jlg} and \citet{hardcastle03jlg} with
independently selected central coordinates.  The coordinates were
optimized so that a fixed radius of 3\,arcsec includes the majority of
the emission associated with the knot and is larger than the PSF in
all observations.  In the cases where the knot is close to the
pointing center this fixed radius slightly overestimates the flux from
the knot including some background, although the majority of this
excess is removed during the background subtraction.  We compared the
X-ray flux of our fixed-radius regions with the flux measured using
regions with radii that were modelled using the PSF and found that the
changes in the light curve reflected the changes in the PSF between
observations.  We therefore use fixed radii regions to eliminate this
effect.  We use annular background regions to account for spatial
variations in the underlying diffuse emission.  We have investigated
the systematic uncertainty due to variation in the surface brightness
within these annular background regions and find that the contribution
to the X-ray flux denisty is negligible (less than 1\%), even in the
worst affected knots.  As AX1A and AX1C are very close together,
manually sized regions were used, adjusted to include as much of the
emission as possible, without including too much emission from the
neighboring knot.  The spectral properties of all the X-ray knots are
shown in Table~\ref{tab:xrayprop}.

\begin{table}
  \centering
    \caption{Spectral Properties of the X-ray Knots}
    \begin{tabular}{lrll}
      \hline
        Knot & Flux Density$^{\dagger}$ & Spectral Index & $N_{H}$ \\
             & 1\,keV (nJy) & $\alpha_{X}$  & ($\times 10^{22}$\,cm$^{-3}$) \\
      \hline
        AX1A & $10.65 \pm 1.64$ & $1.08 \pm 0.04$ & $0.51 \pm 0.02$ \\
        AX1C & $21.43 \pm 2.97$ & $1.06 \pm 0.04$ & $0.52 \pm 0.02$ \\
        AX2  & $ 9.29 \pm 0.54$ & $0.77 \pm 0.07$ & $0.55 \pm 0.04$ \\
        AX2A & $ 3.19 \pm 0.96$ & $0.56^{*}$ & $0.084^{*}$ \\
        AX3  & $ 4.02 \pm 0.39$ & $0.78 \pm 0.16$ & $0.40 \pm 0.07$ \\
        AX4  & $ 4.98 \pm 0.28$ & $0.94 \pm 0.16$ & $0.45 \pm 0.04$ \\
        AX4A & $ 0.30 \pm 0.03$ & $0.48^{*}$ & $0.084^{*}$ \\
        AX5  & $ 6.60 \pm 0.39$ & $0.59 \pm 0.09$ & $0.59 \pm 0.08$ \\
        AX6  & $ 9.39 \pm 0.35$ & $0.51 \pm 0.06$ & $0.59 \pm 0.05$ \\
        BX1  & $ 3.69 \pm 0.27$ & $0.83 \pm 0.11$ & $0.12 \pm 0.02$\\
        BX2  & $19.39 \pm 0.97$ & $0.63 \pm 0.02$ & $0.11 \pm 0.01$ \\
        BX2A & $ 1.44 \pm 0.12$ & $0.58 \pm 0.12$ & $0.084^{*}$ \\
        BX3  & $ 2.02 \pm 0.17$ & $1.40 \pm 0.35$ & $0.15 \pm 0.10$ \\
        BX4  & $ 4.94 \pm 0.30$ & $0.91 \pm 0.06$ & $0.084^{*}$ \\
        BX5  & $ 3.06 \pm 0.23$ & $1.13 \pm 0.15$ & $0.10 \pm 0.03$ \\
        CX1  & $ 2.36 \pm 0.20$ & $1.23 \pm 0.17$ & $0.084^{*}$ \\
        CX1A & $ 0.97 \pm 0.07$ & $2.20^{*}$ & $0.084^{*}$ \\
        CX2  & $ 3.59 \pm 0.37$ & $0.79 \pm 0.06$ & $0.12 \pm 0.04$ \\
        CX3  & $ 0.82 \pm 0.09$ & $0.63^{*}$ & $0.084^{*}$ \\
        CX4  & $ 1.09 \pm 0.18$ & $0.54 \pm 0.20$ & $0.084^{*}$ \\
        EX1  & $ 0.78 \pm 0.14$ & $0.60^{*}$ & $0.084^{*}$ \\
        EX2  & $ 0.58 \pm 0.06$ & $0.60^{*}$ & $0.084^{*}$ \\
        FX1  & $ 1.92 \pm 0.11$ & $0.77 \pm 0.12$ & $0.084^{*}$ \\
        FX1A & $ 0.34 \pm 0.09$ & $0.70^{*}$ & $0.084^{*}$ \\
        FX2  & $ 2.11 \pm 0.25$ & $1.28 \pm 0.27$ & $0.12 \pm 0.09$ \\
        FX3  & $ 1.36 \pm 0.16$ & $1.06 \pm 0.30$ & $0.084^{*}$ \\
        FX5  & $ 0.52 \pm 0.03$ & $1.20^{*}$ & $0.084^{*}$ \\
        FX6  & $ 0.77 \pm 0.08$ & $1.56^{*}$ & $0.084^{*}$ \\
        FX6A & $ 0.47 \pm 0.06$ & $1.20^{*}$ & $0.084^{*}$ \\
        FX7  & $ 1.25 \pm 0.18$ & $1.07 \pm 0.17$ & $0.084^{*}$ \\
        GX1  & $ 1.10 \pm 0.15$ & $0.87 \pm 0.19$ & $0.084^{*}$ \\
        GX2  & $ 1.00 \pm 0.13$ & $0.63 \pm 0.14$ & $0.084^{*}$ \\
        GX3  & $ 1.21 \pm 0.30$ & $1.24 \pm 0.50$ & $0.15 \pm 0.11$ \\
        GX4  & $ 0.61 \pm 0.10$ & $1.20^{*}$ & $0.084^{*}$\\
        GX5  & $ 0.13 \pm 0.03$ & $1.20^{*}$ & $0.084^{*}$\\
      \hline
      \multicolumn{4}{l}{$^{\dagger}$ weighted mean 1\,keV X-ray flux density of the six 100\,ks}\\ 
      \multicolumn{4}{l}{ observations taken in 2007}\\
      \multicolumn{4}{l}{* parameter fixed as too faint for joint fitting; the spectral index is} \\
      \multicolumn{4}{l}{an average of the local fitted indices and the nH is the Galactic value.} \\
   \end{tabular}
  \label{tab:xrayprop}
\end{table}

\subsection{Point Source Contamination}
\label{sec:ps}

\citet{kraft02jlg} investigated whether some of the apparent X-ray
knots could be low mass X-ray binary (LMXB) in Cen A or background
AGN. They simulated point sources using the first of the Chandra X-ray
data sets used in this work (ObsID 0316) to determine whether the
compact knots were point-like enough in their observations to be
confused with LMXBs and found many that are consistent.  They also
considered the radial surface-density distribution of X-ray sources in
Cen A, concluding that they expected $\sim 3$ of the knots or sources
to be X-ray binaries within Cen A unrelated to the jet.

It has been shown (see \citet{fabbiano06jlg} for a review) that a
significant fraction of LMXBs observed in early-type galaxies with
{\it Chandra} are associated with globular clusters (GCs).  It has
recently been confirmed that 41 X-ray point sources in our Cen A data
are associated with GCs within Cen A
\citep{woodley08jlg,jordan07jlg,voss09jlg}.  We examined {\it
  Spitzer}/IRAC 3.6- and 8-$\mu$m IR maps \citep{hardcastle06jlg} to
check for IR counterparts to our jet knots, which would indicate a
coincident globular cluster and found a compact IR source for GX3. We
also checked the GC catalogue by \citet{jordan07jlg} and found a GC
coincident with GX3 only.  Although we do not expect all LMXBs to
reside in GCs, we can rule out any knot that has IR emission as likely
to be a LMXB due to its association with a probable GC.  Since $>$70\%
of all LMXBs identified with GCs lie in the redder GCs, those with a
higher (near-solar) metal abundance
\citep[e.g.,][]{woodley08jlg,pb09}, the latter would actually be more
readily detectable in the near-IR {\it Spitzer}/IRAC observations than
in ground-based optical images.

The X-ray knots AX2A and SX1 are compact X-ray sources with no compact
or diffuse radio emission, and they lie outside the boundaries of the
detected radio jet and counterjet.  However, the X-ray flux
variability of AX2A is substantially different from that of SX1; it
was undetected until 2007 when it flared to 3\,nJy, and it has not
varied significantly since. AX2A may therefore be a genuine new X-ray
knot rather than a LMXB so is considered further in
Section~\ref{sec:dis}.

Contamination from background AGN is highly unlikely; these would
appear as point sources, possibly with optical counterparts and
generally with flatter X-ray spectra \citep[typical unabsorbed X-ray
  AGN spectra have spectral indices $1.09\pm0.08$, ][]{mainieri07jlg}.
We have also calculated the number of AGN we expect in the jet using
the background $log N - log S$ method described by
\citet{moretti03jlg} and find only a 33\% chance of finding an AGN in
our jet.

\subsection{Emission Mechanism}
\label{sec:em}

Using the three frequencies of radio data observed in 2007, we fitted
a synchrotron model to the radio emission from the inner A-group
knots, which has allowed us to predict the X-ray emission we would
expect from synchrotron self-compton emission (SSC) and from the
inverse Compton scattering of the cosmic microwave background (iC/CMB)
and of the galaxies optical star light (iC/SL).  We used the sizes
measured by \citet{tingay09jlg} to estimate the emitting volume of the
stationary knots, which appear to have compact cores, combining the
volumes of the substructures in the cases of A1A and A2A.  We used the
radio fluxes measured from radio maps matched to the resolution of the
22\,GHz data ($1.80 \times 0.40$\,arcsec) and the weighted mean,
1\,keV X-ray flux density from the six 2007 X-ray observations.  At
equipartition magnetic field strengths, the observed X-ray emission is
much greater than the predicted X-ray flux density, which is dominated
by SSC for the stationary knots, A1A, A1C, and A2A.  If we assume the
SSC model is dominant in the X-ray regime, we find that the magnetic
field strengths required for the observed X-ray emission, for the
stationary knots, are a factor of 500 -- 600 weaker than the
equipartition values.  This is also true for the knots that are not
detected by \citet{tingay09jlg} and that are unresolved in our data; a
limit on the sizes was used to find the limits on the equipartition
magnetic field strengths and the internal energies and pressures of
the knots.  Table~\ref{tab:synch} shows the radius of the emitting
volume, the radio and 1\,keV X-ray flux densities, the equipartition
magnetic field strengths, $B_{eq}$, and the required magnetic field
stength for SSC dominated X-ray emission, $B_{SSC}$, of the A-group of
knots.

In other features, such as the hotspots in FR\,II radio galaxies, the
magnetic field strengths required for the observed X-ray are only
slightly less than the equipartition values \citep[factors of 3 --
  5][]{hardcastle04jlg,kataoka05jlg}.  The much larger departure from
equipartition required for iC to be significant in the Cen A knots,
combined with the steepness of the spectral indices for these knots
($\alpha_{X} > \alpha_{iC}\sim0.5$) suggest that iC emission is not
significant in the X-ray for the majority of the jet knots.  We
therefore assume that the X-rays from the knots are synchrotron
emission in the remainder of this work.

Using the equipartition magnetic field strengths, we were able to
estimate the total energy density of the knots and find that the
internal pressures of these knots are of the order of 1\,nPa, which is
much higher than the pressure in the surrounding diffuse material.
This is also evident from the higher surface brightness of the knots,
which is directly related to the internal energy of the knot material.

\begin{table*}
  \centering
    \caption{Emission model parameters for the inner A-group knots}
    \begin{tabular}{lrrrrrrrrr}
      \hline
        Knot & Radius & \multicolumn{4}{c}{Flux Density} & X-ray lifetime & \multicolumn{2}{c}{Magnetic Field Strength} & Pressure\\
        & & 4.8\,GHz & 8.4\,GHz & 22\,GHz & 1\,keV X-ray & $\tau_{1keV}$ & $B_{eq}$ & $B_{iC}$ & $P_{int}$\\
        & (pc) & (mJy) & (mJy) & (mJy) & (nJy) & (yrs) & (nT) & (nT) & (nPa)\\
      \hline
        A1A/AX1A & 2.017    & $20.10 \pm 4.26$ & $12.27 \pm 2.04$ & $ 5.32 \pm 0.98$ & $10.65 \pm 1.64$ & 5.61 & 69.3 & 0.119 & 0.955\\
        A1B      & $<$6.586 & $46.92 \pm 5.58$ & $29.44 \pm 2.67$ & $13.83 \pm 1.28$ & $<19.38$ & 17.63 & 32.3 & $\ge0.125$ & 0.937\\
        A1C/AX1C & 2.293    & $41.45 \pm 5.66$ & $25.33 \pm 2.70$ & $12.32 \pm 1.30$ & $21.43 \pm 2.97$ &  4.86 & 76.3 & 0.135 & 1.546\\
        A2A/AX2  & 2.727    & $37.13 \pm 5.82$ & $15.57 \pm 2.78$ & $ 6.32 \pm 1.33$ & $ 9.29 \pm 0.54$ &  7.46 & 57.3 & 0.119 & 0.871\\
        B1A/BX2  & 15.300   & $ 2.64 \pm 0.64$ & $ 0.93 \pm 0.61$ & - & $19.39 \pm 0.97$ & 30.89 & 22.2  & 0.038 & 0.131 \\
      \hline
   \end{tabular}
  \label{tab:synch}
\end{table*}


\subsection{Proper Motions}
\label{sec:pm}

The results of \citet{hardcastle03jlg} were based on only the first
two epochs of radio data.  With these data, they were able to
establish the bulk flow speed of the jet ($\sim 0.5$\,c) and also
demonstrated that some of the knots move along the jet (A1B, A2, A3B,
and A4) while others were consistent with being stationary (A1A, A1C,
A2A, A3A, A5A, B1A, SJ1, SJ2, SJ3, S2A, and S2B).  With our multi-epoch
data we can improve on the accuracy with which the proper motions are
measured.  We used maps with a matched resolution of $0.80 \times
0.20$\,arcsec for these measurements.

Our approach to fitting speeds was to use a modified version of the
shift-and-fit method of \citet{walker97jlg}, as used in a simpler form
by \citet{hardcastle03jlg}.  As we have more than two maps we
attempted to fit a velocity vector, consisting of an angular speed and
direction, to each knot. (More complex models are not justified by the
quality of the data.) To use the shift-and-fit method we selected a
reference image at a particular epoch. For a given trial value of the
angular velocity vector the appropriate part of this image was then
shifted (using a bicubic polynomial interpolation) to the position
implied for all the other epochs, the difference of the two images was
formed, and the contribution to $\chi^2$ was calculated using
estimates of the local on-source noise in both maps. The total
$\chi^2$ over all non-reference images was minimized using a
Markov-Chain Monte Carlo algorithm (briefly described by Croston et
al. 2008) which allows the efficient exploration of parameter space. A
Jefferys (scale-invariant) prior was used for the magnitude of the
angular velocity vector to avoid bias towards large values. In
principle this algorithm also allows an efficient determination of the
uncertainties (formally the credible intervals) on the fitted
parameters. However, we found that these errors were dominated by the
systematic uncertainties due to the choice of reference image; in weak
knots a fortuitous distortion in the reference image can give the
appearance of a proper motion that is not actually present. To remedy
this we carried out the fits for a given knot using each of our seven
radio images in turn as the reference image. Only knots in which
consistent, non-zero motions are detected for all choices of the
reference image are considered to be moving. In these cases our best
estimate of the speed of the motion is the median of the Bayesian
estimates of the angular speed for each choice of reference image, and
the range of speeds returned under different choices of the reference
image gives us an estimate of the systematic uncertainties in the
result.  Where the velocities are inconsistent, we have taken the
upper limit to be the largest velocity in this range.

\begin{figure}
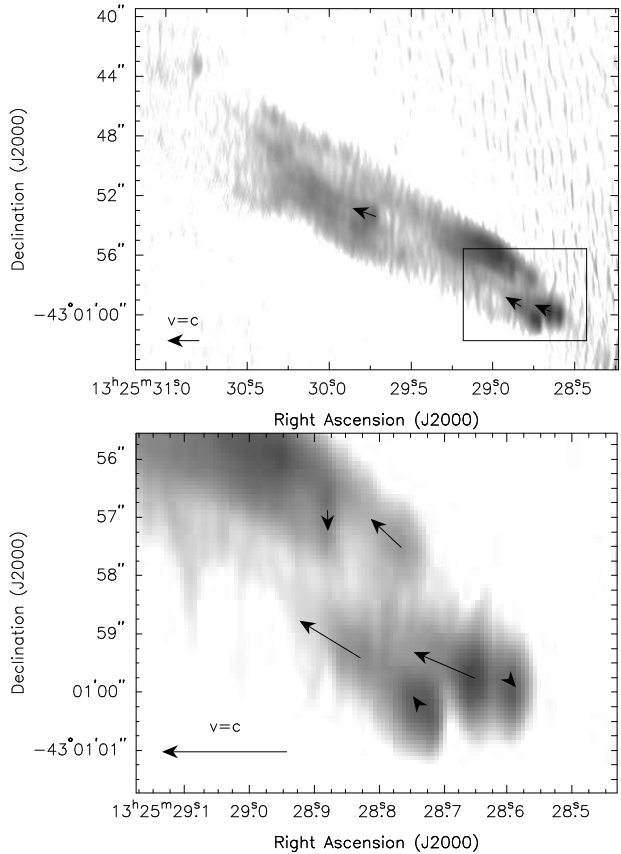

\includegraphics[width=0.45\textwidth]{vel-vec.ps}
\includegraphics[width=0.45\textwidth]{vel-vec-zoom.ps}
\caption{Velocity vectors for the moving knots in Cen A, with the composite 8.4\,GHz radio map convolved to a resolution of $0.8\times0.2$\,arcsec showing the well-established velocities for the A-group of radio knots (left panel) and the mean velocities of the A1 and A2 knots is shown in the right panel.  The region included in the right image, is shown with the box on the left image.  Black corresponds to 0.1\,Jy and white corresponds to 0.1\,mJy.   A velocity of $1c$ is shown in the bottom left corner of these images.}
\label{fig:vectors}
\end{figure}

This approach detected apparent motions in 6 of the radio knots as
well as in the diffuse material downstream of the A2A knot (the
regions downstream of A2A are labelled A2B, A2C and A2D in the
following sections).  To check that these motions are sensible, we
verified the motion visually.  We found that two of these knots, A1C
and A2A are actually stationary and we attribute the detected proper
motion to the evolution of the knot; A1C appears to grow downstream
while A2A's front edge is stationary with diffuse material appearing
to break off and move downstream from this knot towards A2B, A2C and
A2D [all consistent with \citet{hardcastle03jlg}].  The visual checks
also rule out the apparent motion in SJ1 as its proximity to the
bright core means its shape is affected by artefacts.  We are left
with three knots moving in the jet: A1B, A1E, and A3B (their
velocities are plotted as vectors in Fig.~\ref{fig:vectors} and shown
in Table~\ref{tab:summaryjlg} with the limits for the stationary
knots).  This approach is more robust that that of
\citet{hardcastle03jlg} as it considers all seven epochs of our radio
data, reducing the errors on the proper motions of the moving knots,
and we have constrained the speeds of the other knots that had no
previously detected proper motions.

It is worth noting here that A1B and A1E have no X-ray emission
associated with them while the region A3B can be described as
consisting of three sub-regions in the radio with a diffuse X-ray
counterpart, possibly breaking any correlation between compact,
radio-only knots and proper motions.

\begin{figure}
\includegraphics[width=0.32\textwidth,angle=270]{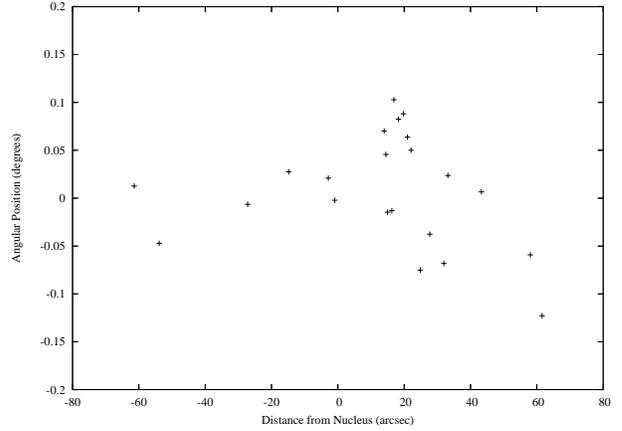}
\caption{The offset between the position angle of the jet in Cen A ($54^\circ$)  and the transverse position of the radio knots as a function of the projected distance from the nucleus for the knots in the jet and counterjet (negative distances).  The knots do not appear to lie at the PA $54^\circ$ for the entire length of the jet.}
\label{fig:swing}
\end{figure}

We have also determined the directions of travel for the
well-established moving knots (Table~\ref{tab:summaryjlg}).  The axis
of the inner, hundred-parsec-scale jet has a position angle (PA) of
$54.1^\circ$ east of north from the core and its extrapolation
provides a good estimate of the axis of much of the outer jet.  The
moving knots all travel in directions eastward of this axis.  They
also all belong to the A-group of knots and, on closer inspection of
this section of the jet, we find that it also deviates eastward to
greater PAs.  The jet axis in this region has a PA of $62.3^\circ$.
Two of our three knots move in a direction consistent with this,
within $3\sigma$ errors.  If the jet motion was purely conical,
expanding directly away from the core, the motion of the knots should
be radial, but we find that the directions of motion do not match the
knot PA's.  The ridge line through the A-group knots may actually
follow the regions of highest radio surface brightness, swinging from
north of PA $54.1^\circ$ at A2 to south of it at A3 and A4.  In
Figure~\ref{fig:swing}, deviations of the knot PA's from $54.1^\circ$
are plotted against distance from the core, clearly showing this swing
in the ridge line.  We can only conclude that the fluid flow along the
jet is neither laminar nor in a straight line away from the core,
consistent with a complex flow.

The median speeds of the remaining radio knots are generally smaller
than the speed of the slowest knot with a definite detection, A1E,
($v/c \sim 0.34$), but in most cases the upper limits exceed $0.5c$,
allowing the knots to be moving.  We require better data to establish
whether this is the case, so for the remainder of this work, we have
considered those with well established velocities to be moving, those
with low median speeds ($<0.2\,c$), and low upper limits to be
stationary and we classify the others as inconclusive.  In
Table~\ref{tab:summaryjlg} these classifications are indicated by Y,
N, and I respectively.  We discuss the association between the motions
of the knots and their other properties in
Section~\ref{sec:dis-moving}.

\subsection{Flux Variability}
\label{sec:flux-var}

Another important property that can be measured from our multi-epoch
radio and X-ray data is flux variability.  Utilizing the multi-epoch
8.4\,GHz radio data, we have been able to monitor the radio flux
variability over the last 17 years.  We used radio maps of matched
resolution ($0.80 \times 0.20$\,arcsec) for this analysis to eliminate
any flux variation due to varying PSF. The initial light curves for
the radio knots showed a systematic variation of up to $\pm 10\%$
common to all knots, which we attributed to differences in the flux
calibration, so we normalized the radio fluxes using a weighted mean
of the brightest compact knots (A1A, A1B, A1C, A2A, A3B, and A4).  We
chose these knots as the others are weaker and/or more diffuse, and
would therefore introduce large uncertainties in our weighted mean
value. We also excluded SJ1, which shows strong variation since 2004.
The core was not used in this normalization as it is known to vary.
The normalized fluxes are shown in Table~\ref{tab:radiofluxjlg}.  We
fitted a constant to these radio light curves and minimized the
$\chi^{2}$ to determine whether the radio fluxes are at all variable,
with a reduced $\chi^{2} < 2.80$ being the threshold for a constant
radio light curve (99\% confidence for 6 d.o.f.).  We detect radio
variability at this confidence level in 9 of the 19 radio knots
(49.4\%).  The light curves of these varying knots are shown in
Appendix A; Figures~\ref{fig:lc1} and~\ref{fig:lc2} show the light
curves for those radio knots with X-ray counterparts and
Figures~\ref{fig:rlc1} and~\ref{fig:rlc2} show the radio and X-ray
light curves for the radio-only knots.  The most noticeable variation
is in the counterjet knot SJ1, which has increased in flux by a factor
of 3 since 1991.  Three of these radio variable knots show
fluctuations on yearly time scales (B2, S1 and S2A), while the
remaining four split into two increasing (A1C and SJ1) and two
decreasing (A1B and A1D) gradually over the 17 years.

We also considered variability in the 4.8\,GHz data but as we have
only 4 observations over 24 years at irregular intervals, we cannot
comment on any short-term variability.  These light curves are all
broadly consistent with the 8.4\,GHz light curves, an example of which
is shown in Appendix A, Figures~\ref{fig:rlc1} and~\ref{fig:rlc2}.
Accepting a reduced $\chi^{2} < 3.78$ as a constant light curve (99\%
confidence for 3 d.o.f.) we find 6 radio knots with some degree of
variability at 4.8\,GHz, half of which are also variable in the
8.4\,GHz data with the detected 8.4\,GHz variable knots, which is not
to say that those which are apparently constant do not agree with the
8.4\,GHz variability.  This is particularly evident in the SJ1 data,
as only two observations overlap with the time baseline of the
8.4\,GHz data; these data could be interpreted as decreasing while the
4.8\,GHz data increases, but there are too few data to draw any
conclusion from the 4.8\,GHz light curves.  The dynamic ranges of
these 4.8\,GHz data are much lower than our 8.4\,GHz data so they are
subject to much larger systematic errors.

We also detected long- and short-term X-ray variability.  Combined
with the radio variability, this allows us to search for any changes
in beaming or particle acceleration properties.  We carried out a
joint fit to all 10 X-ray data sets for all the X-ray knots, fitting a
single photon index and column density for each knot, but allowing the
normalizations to vary in order for any variations in the flux to be
detected.  The normalizations were converted to 1\,keV flux densities
so the light curves could be plotted including 1$\sigma$ errors.  We
then fitted a constant to the light curves, minimizing the $\chi^{2}$
to find the best fit.  We were able to carry out a joint fit for 24/40
of the X-ray knots, 22 of which have a $\chi^{2}_{red} < 1.10$.  For
the remaining knots where there were not enough counts for spectral
fitting, we firstly determined whether there was a $3\sigma$ detection
of the knot considering each observation separately.  Where the knot
was detected, we fixed the photon index to the average photon index of
the nearby knots.  The flux was then determined from the background
subtracted counts and the model count rate from \textsc{xspec}.  Where
the knot was not detected, the $3\sigma$ limit was calculated.

Knots with a $\chi^{2}_{red} > 2.41$ (99\% confidence for 9\,d.o.f.)
are considered variable in the X-ray.  We find that 5 X-ray knots vary
(12.5\%) in addition to AX2A and SX1 which are candidate LMXBs (see
Section~\ref{sec:ps}), 3 of which have varying radio counterparts.
The light curves for the X-ray varying knots are shown in Appendix A;
Figure~\ref{fig:xlc} shows the X-ray only knots while those that have
radio counterparts are shown in Figures~\ref{fig:lc1} and
~\ref{fig:lc2}.

To test whether the variability behavior is consistent with, or
different from that of the non-jet point sources, we also extracted
spectra for the off-jet point sources.  We manually checked the
results of {\it celldetect} to remove detections of image artifacts
and the jet knots before running the same fitting using annular
background regions.  We detected 423 point sources, only 183 of which
are detectable in all observations due to changes in the pointing and
roll angle of {\it Chandra} for each observation causing slightly
different regions of the sky being imaged.  Although the PSF changes
across the image for each observation, these point sources all lie
within 5.5\,arcmin of the core, so the changes in the PSF of each knot
is not significant.  However, the changes in PSF between each
observation is more significant and is reflected in the light-curves
causing an apparent flux variability.  To remove this affect from the
light-curves, we have used regions with a fixed radius of 3\,arcsec
for these point sources.  We fitted a constant to the X-ray light
curves of these point sources and found 105 show some degree of
variability with 99\% confidence ($56.8 \pm 10.3$\% of the point
sources), a factor of 3 more than the jet X-ray knot population.  In
41 instances, the background annulus contained zero counts so these
were re-extracted with a larger background annulus to determine a
limit.  If we consider the point sources within 3.5\,arcmin of the
core, which limits us to the length of the jet from the core, we find
that 76/141 point sources vary ($53.9 \pm 8.7$\%) and if we reduce the
sample further to include only those on the East of the image, so
those within $45^{\circ}$ of the jet PA, we find that 33/71 point
sources vary ($46.5 \pm 5.7$\%).  We can see that the effect of the
PSF is not significant in these samples and we can conclude that we
are looking at a group of different objects in the jet and not just
coincidentally positioned X-ray binaries in most cases.

\subsection{Polarimetry}
\label{sec:pol}

\begin{figure*}\hskip 2cm
\includegraphics[width=0.80\textwidth]{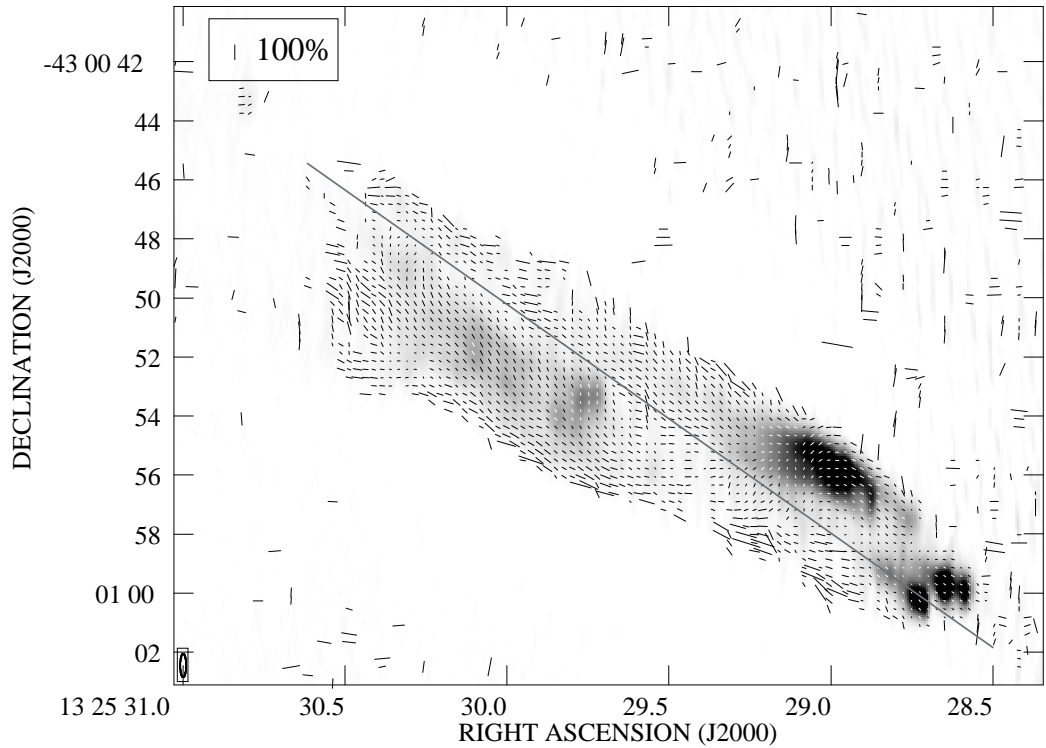}
\caption{Radio polarization in the inner jet of Cen A showing the direction of the magnetic field vectors on a composite, Stokes I, 8.4\,GHz radio map.  The radio image is in the range 0.1 -- 0.01 Jy and a vector 0.56\,arcsec long represents 100\%.  The solid grey line highlights a possible axis of null polarization discussed in Section~\ref{sec:pol}.}
\label{fig:magjlg}
\end{figure*}

If there is compression/rarefaction of the plasma in the knots, we
would expect changes in the polarization, as the magnetic field is
assumed to be frozen into the plasma.  We can therefore use the
polarization data to investigate any link between the activity of the
knots with physical changes in the plasma.  The co-added, matched
resolution, Q and U images at 8.4\,GHz were used to make a deep
magnetic field vector map shown in Figure~\ref{fig:magjlg}.  We made
the individual Stokes Q and U maps using the \textsc{aips} task {\it
  imagr}.  There are artifacts around the core for all epochs, which
we attribute to the limited accuracy of the correction calculated for
the ``leakage'' terms determined by {\it pcal}; only the 2003 data set
is unusable due to higher noise in these Stokes Q and U images.  The
overall direction of the magnetic field is down the jet, consistent
with what was found by \citet{hardcastle03jlg} with no obvious change
in the knots.

\begin{figure*}\hskip 2cm
\includegraphics[width=0.80\textwidth]{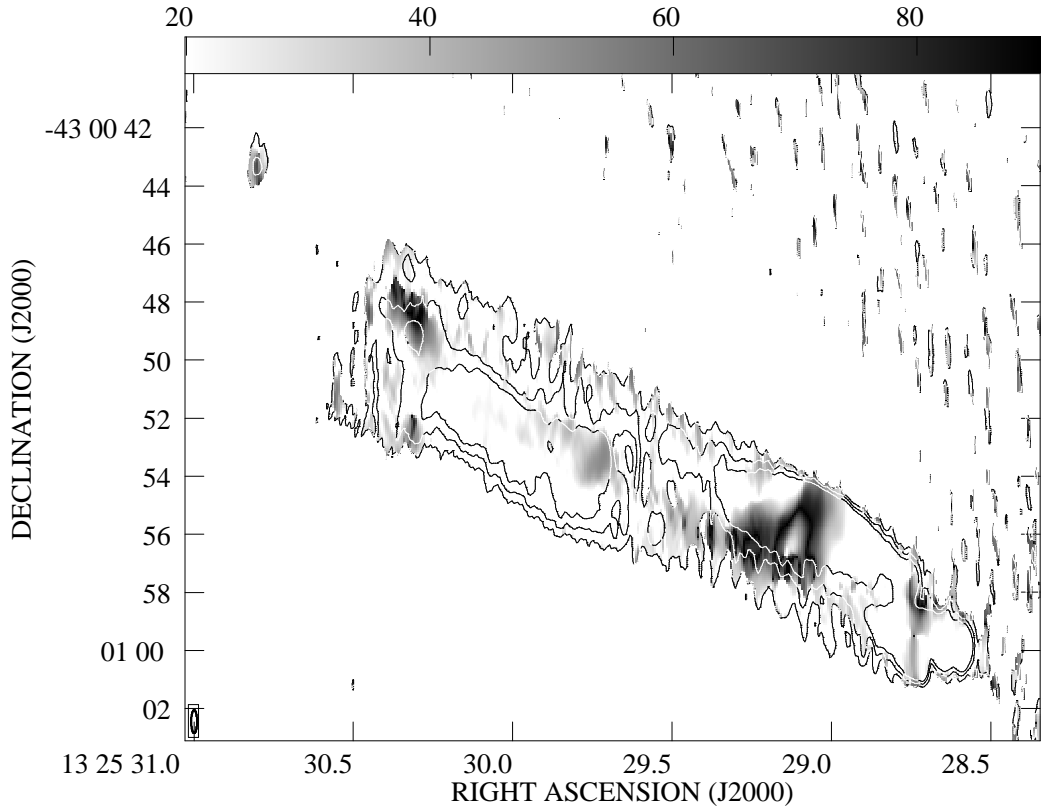}
\caption{Greyscale map of the inner jet of Cen A showing the difference between the  direction of the magnetic field vectors and the jet PA (51.4$^{\circ}$) with total intensity contours at 3, 8 and 16 times the rms of the composite, Stokes I 8.4\,GHz radio map, $8.23\times10^{-5}$\,Jy beam$^{-1}$.  The greyscale image is in the range 20 -- 90 degrees to emphasise structure in which the B-vectors are substantially misaligned with the jet.}
\label{fig:PAjlg}
\end{figure*}

We detected a $3\sigma$ variation in residual maps of the Q and U
Stokes parameters (2007-2002) in the A1 and A2 groups.  The diffuse
emission farther down the jet also showed evidence of low level
variability, but as this was uniform across the region it was
attributed to differences in the flux calibration.  We proceeded to
measure the fluxes in Q, U and I for these data, normalizing as
described in Section~\ref{sec:flux-var}.  Comparing the total
intensity (I), the angle of polarization ($\theta =
\frac{1}{2}\arctan(U/Q)$, in this work we use the $Q/U$ ratio as an
approximation to this relationship) and the degree of polarization ($p
= \sqrt{Q^{2}+U^{2}}/I$) allowed us to identify variations due to a
change in the total intensity, the polarization intensity, the
polarization position angle or those due to the proper motion of the
knot.

As previous described, the majority of knots show only a low level
variation in total intensity; however 4/19 show changes in the degree
of polarization only: B2, SJ1, S1 and S2B (Appendix A,
Figure~\ref{fig:poldeg}), and 6/19 show changes in the angle of
polarization (Appendix A, Figure~\ref{fig:polang}) including two knots
which show changes in both.  We have to consider that one of these
knots is moving, A1B, so the observed changes in polarization could be
due to this movement.  Excluding this moving knot, the number of knots
varying only in the angle of polarization is unchanged, and we have
one knot, A1A, which is changing in both.  These results are compared
with the total radio and X-ray flux variations in
Section~\ref{sec:summary}.

The A2 diffuse knot (A2B, A2C, and A2D) shows a change in the $Q/U$
ratio indicating a change in the polarization angle across the region;
however, there are no X-ray counterparts for any of these sub-regions
and their radio spectral indices are consistent ($\alpha_{4.8}^{8.4} =
1.13 \pm 0.10$, $1.01 \pm 0.16$ and $1.22 \pm 0.27$ respectively).
Only the central section, A2C, varies in total radio intensity but the
entire region appears to be moving downstream away from A2A so this
change in polarization may be due to the motion of this diffuse
material.  The detected variation in the A1 group knots A1A, A1B and
A1C, cannot be attributed to motion of the knots, as only A1B is
moving.  In this instance, the observed polarization variability may
be explained by compression and rarefaction of the knot material.  We
do not see any perpendicular field structure across any of the knots,
which might be expected in a local shock model; however, this might be
masked by complicated jet polarization structure.  {\edit However, we
  do see systematic misalignments between the jet PA and the magnetic
  field associated with some jet features, notably A2 and A3B, which
  are highlighted in Figure~\ref{fig:PAjlg} which shows the difference
  between the polarization angle and the PA of the jet.}
\citet{clarke92jlg} found only a modest rotation measure (RM) in the
inner lobes and jet, and only a slight change in the Faraday corrected
magnetic field vectors in their 6cm (4.8\,GHz) radio data, so we do
not expect the effect of RM to to be significant.  We will discuss the
effect of various models on the polarization of the knots further in
Section~\ref{sec:dis}.

It is interesting to note that there is an apparent null in the
polarisation which crosses the diffuse material of the A-group region,
indicated on Figure~\ref{fig:magjlg} by a grey solid line.  This line
extends to the core directly through the inner hundred-parsec-scale
jet suggesting that it is associated with magnetic fields originating
in or close to the core.  This line also splits the bright A1 and A2
complexes from the A3 and A4 complexes.  This could plausibly be a
result of a helical jet field structure.

\subsection{Spectral Indices}
\label{sec:index}

\citet{hardcastle07jlg} and \citet{worrall08jlg} investigated the
X-ray spectrum of the jet in the longitudinal and transverse
directions, respectively.  \citet{hardcastle07jlg} showed that the
inner jet is dominated by knots, consistent with local particle
acceleration at shocks, while further down the jet steeper-spectrum
diffuse X-ray emission is more dominant.  \citet{worrall08jlg} found
that in the knotty region beyond the A1 and A2 complexes and within 66
arcsec of the core, the weighted X-ray spectrum of knots closer to the
jets axis (the `inner spine') is harder than that further off axis
(the `inner sheath') ($\Delta \Gamma = 0.31 \pm 0.07$).  This was
interpreted as evidence that the jet speed is higher closer to the
axis, with more kinetic energy available for producing a harder X-ray
spectrum.

\begin{figure*}
\includegraphics[width=0.95\textwidth]{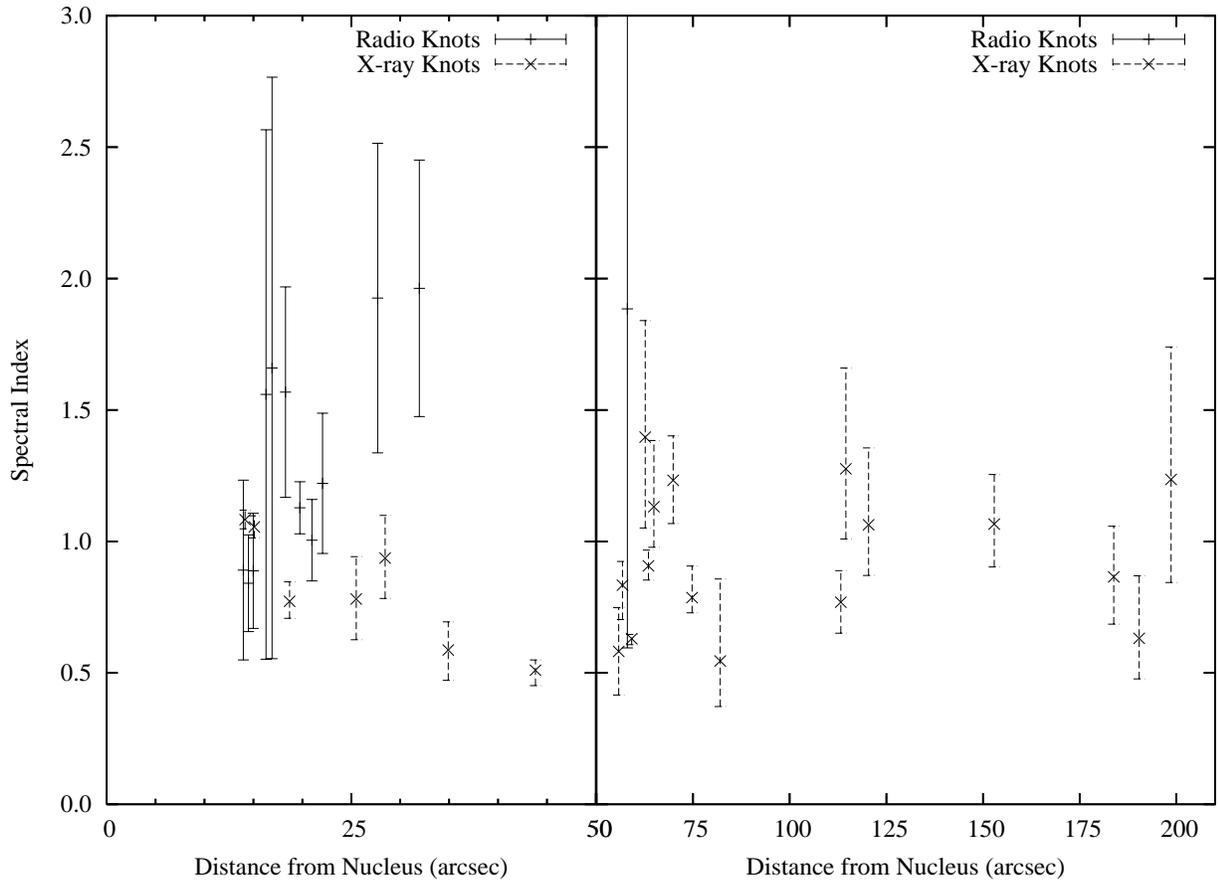}
\caption{The radio spectral index, $\alpha_{4.8}^{8.4}$ (solid lines) and X-ray spectral index, $\alpha_{X}$ (dashed lines) of knots in the jet of Cen A as a function of distance from the core.  The left panel shows only the inner knots (up to 50\,arcsec) and the right panel shows the rest of the jet knots.}
\label{fig:xrdist}
\end{figure*}

\begin{figure*}
\includegraphics[width=0.35\textwidth,angle=270]{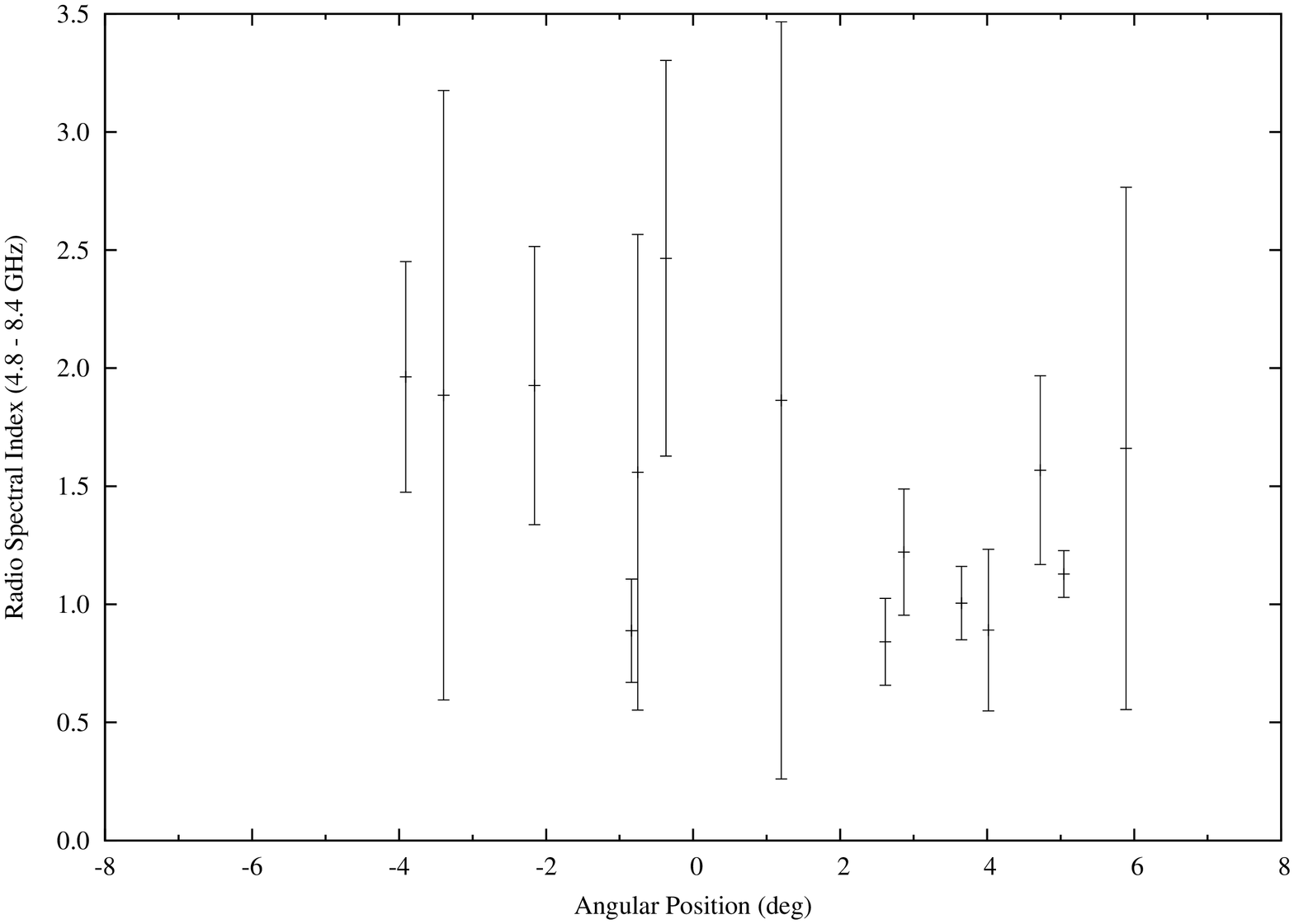}
\includegraphics[width=0.35\textwidth,angle=270]{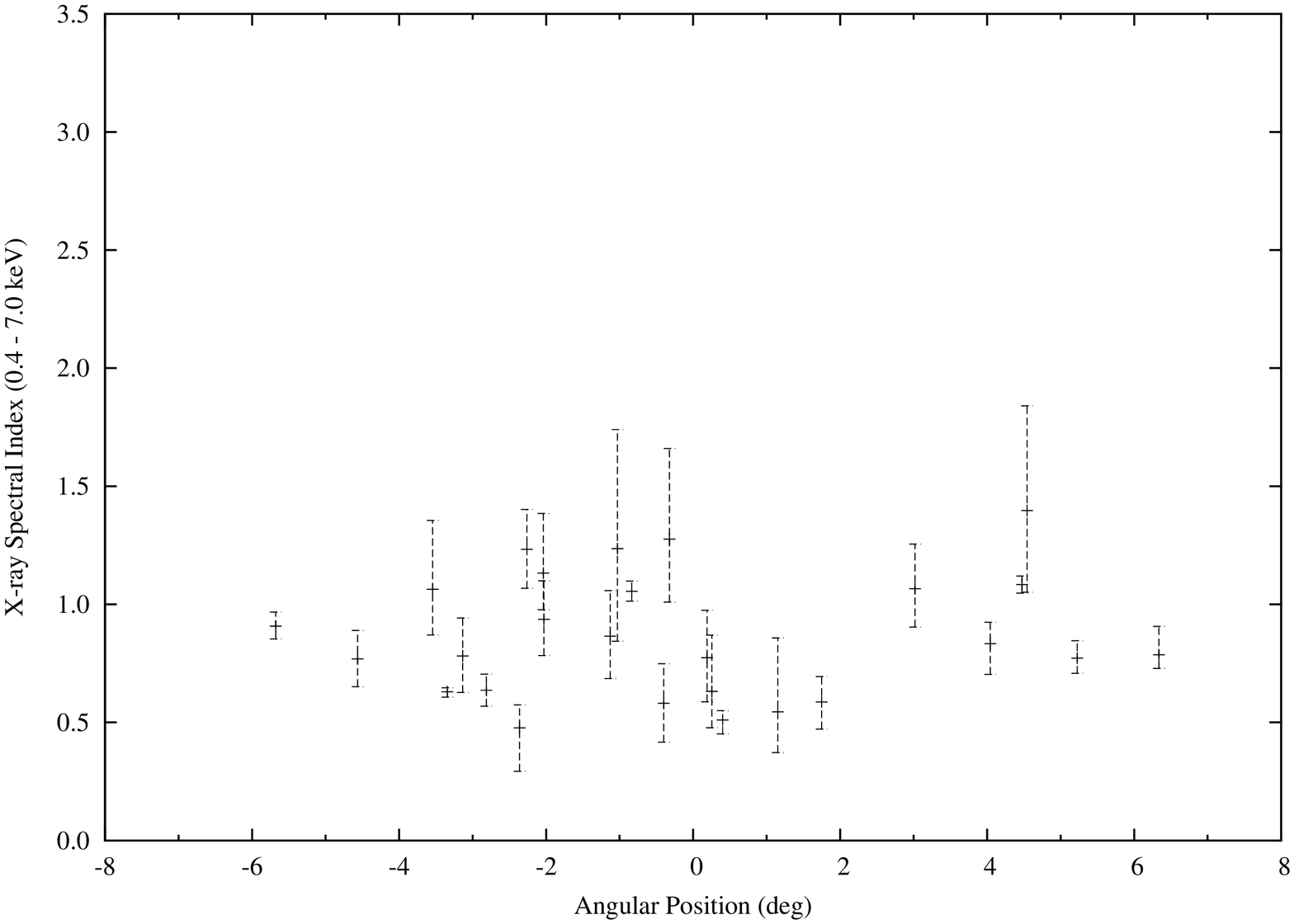}
\caption{The radio spectral index, $\alpha_{4.8}^{8.4}$ (left panel, solid lines) and X-ray spectral index, $\alpha_{X}$ (right panel, dashed lines) of knots in the jet of Cen A as a function of offset between the angular position of the knot and the jet position angle of $54.1^\circ$.}
\label{fig:xrang}
\end{figure*}

Here we compare the radio and X-ray spectral indices of individual
knots with their longitudinal and transverse positions.  We have
measured the radio spectral indices for all of the radio knots;
however, only 14 have well established indices.  In the X-ray, we have
fitted spectral indices to 26 of the X-ray knots.  We compared the
X-ray spectral indices ($\alpha_{X}=\Gamma-1$) of the jet-side knots,
irrespective of whether they have counterparts, as a function of
distance from the core (Figure~\ref{fig:xrdist}), and of the offset
between the angular position of the knot and the jet PA of
$54.1^{\circ}$ (Figure~\ref{fig:xrang}).  

All the X-ray spectral indices were consistent with those determined
by \citet{hardcastle07jlg} and consistent with synchrotron emission
with $\alpha_{X}>0.5$; the distances from the core of each knot were
also consistent owing to the knot selection process.  When we compared
the measured spectral indices of all the radio and X-ray knots to the
offset between the jet PA of $54.1^\circ$ and the angular position of
the knot (Figure~\ref{fig:xrang}), we find a continuous distribution
of X-ray spectral indices with no statistically significant
correlation according to both a KS and a Wilcoxon-Mann-Whitney test,
although these test do not take account of the errors on our spectral
indices.  We also considered the knots in each of the regions defined
by \citet{hardcastle07jlg} and find no correlation in these regions
either.  These comparisons will be discussed further in
Section~\ref{sec:spine}.  

We have also found the weighted mean of the spectral indices for all
of the knots within the regions defined by \citet{worrall08jlg} as the
`inner-spine' and `inner-sheath',{\edit $\alpha_{X}=0.62\pm0.01$ and
  $\alpha_{X}=0.89\pm0.05$ respectively,} and agree with the findings
of Worrall et al. of a harder weighted X-ray spectrum in the spine
than in the sheath over the same length of jet (although $N_{\rm H}$
changes by a factor of 3, the absorption is fitted so does not affect
the spectral indices).  Within these regions, we have fitted spectral
indices for 6/7 of those in the inner-spine; AX3, AX4, AX5, AX6, BX2,
and BX5, and for 3/4 or those in the inner-sheath; BX1, BX3 and BX4.
Those that are not fitted; BX2a in the spine and CX1a in the sheath,
are very faint knots and as discussed in Section~\ref{sec:ps}.  The
interpretation of these measurements is discussed in
Section~\ref{sec:spine}.

We also calculated the radio spectral indices between 4.8 and
8.4\,GHz for all the radio knots utilizing our multi-frequency 2007
radio data.  Due to the low dynamic range of the 22.5\,GHz map, we
cannot measure fluxes for many of the knots; however, for those which
are resolved, the spectral index for 4.8 -- 22.5\,GHz, $\alpha_{4.8}^{22.5}$,
is consistent with the 4.8 -- 8.4\,GHz spectral index
($\alpha_{4.8}^{8.4}$).  The errors on these spectral index measurements
are dominated by the noise in the images.  These are listed in
Table~\ref{tab:featuresjlg} with the radio and X-ray flux densities,
X-ray photon indices and X-ray/radio ratios, which are discussed in
Section~\ref{sec:counters}.

\subsection{Counterparts}
\label{sec:counters}

We have identified 13 knots that are detected in both radio and X-ray:
A1A and AX1A, A1C and AX1C, A2A and AX2, A3A and AX3, A3B and AX4, A5A
and AX5, A6A and AX6, B1A and BX2, B2 and BX4, SJ3 and SJX1B, S1 and
SX1A, S2A and SX2A, and S2B and SX2B.  Each of the knots in these
pairs have the same coordinates with no significant offsets; we do not
consider knots with possible offset counterparts as paired, given the
discussion of offsets in \citet{hardcastle03jlg}, {\edit who argued
  that the apparent offsets between the radio and X-ray knots in more
  distant radio galaxies are a result of an inability to resolve faint
  aligned radio-knot counterparts from bright downstream diffuse
  emission.}

With the radio and X-ray data for these knots, we were able to measure
the ratio of the 1\,keV X-ray flux density and the 8.4\,GHz radio flux
density and use the median of these X-ray/radio flux density ratios to
determine whether those without detected counterparts are truly
without counterparts or whether the counterpart is too faint to be
detected.  Using the median value of the X-ray/radio flux density
ratios, we predicted the flux density of the missing counterparts for
the radio-only and X-ray-only knots.  The median value of $1.01 \times
10^{-6}$ is used rather than the mean, as the distribution of these
X-ray/radio flux density ratios is not Gaussian. The measured
X-ray/radio flux density ratios range from $0.07 \times 10^{-6}$ to
$9.44 \times 10^{-6}$.  These predicted flux densities assume that all
the knots have the same spectrum with consistent X-ray/radio flux
density ratios.  By comparing these predictions to the measured flux
densities we determined whether the absent counterpart can be
detected.

Out of the 6 radio-only knots, we find that all except two radio
knots, A1D and A1E, should have detectable X-ray counterparts using
the median X-ray/radio flux density ratio value; however, at the lower
limit, only SJ1 should have a detectable X-ray counterpart.
Unfortunately, SJ1 is located only 1 arcsec ($\sim 17$\,pc) from the
core, so in the X-ray, the emission from the knot is contaminated by
the bright core.  In {\it Chandra} HRC observations taken in 1999
\citep{kraft00jlg}, SJ1's X-ray counterpart is still unresolved from
the nucleus despite the slightly higher spatial resolution.  As 7/9
radio knots have detectable but unseen X-ray counterparts when the
median ratio is assumed, they probably have steeper spectra than those
that have detected counterparts suggesting a genuine difference in
their particle acceleration properties.

When we invert this rationale and consider the X-ray-only knots, we
find that 9/27 X-ray knots would have detectable radio counterparts at
the median X-ray/radio flux density ratio and, as they are not
seen, they are likely to have flatter ratios than those that do have
detected counterparts.  However, at the limits of the range of
measured ratios, all these knots are detectable at the lower limit and
not-detectable at the upper limit.

\begin{figure}
\includegraphics[width=0.46\textwidth]{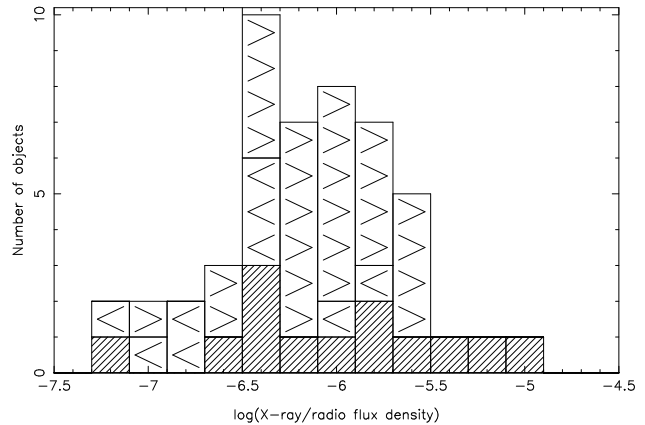}
\caption{A histogram of the X-ray/radio flux density ratios for the paired knots (hashed region), the lower limits on the X-ray/radio flux density ratio for the X-ray-only knots ($>$), and the upper limits on the X-ray/radio flux density ratio for the radio-only knots ($<$) in Cen A.}
\label{fig:hist}
\end{figure}

If we reverse this argument and consider the X-ray/radio flux density
ratio values we would measure if the missing counterpart were at the
limit of the noise in the image, we can obtain upper limits on the
X-ray/radio flux density ratio for the radio-only knots and lower
limits for the X-ray-only knots.  In Figure~\ref{fig:hist}, we show
histograms of the X-ray/radio flux density ratios for the three
populations.  As the radio-only and X-ray-only knots give us limits,
we find that the peaks of all three groups of knots are consistent, so
we can not rule out a single population; however, we will continue to
discuss the knots in three groups and accept that many of the knots
may have the same production mechanisms.

The properties of these three populations of knots are summarized in
Section~\ref{sec:summary}; in Section~\ref{sec:dis} we shall discuss
current knot formation and particle acceleration models and how they
explain the differences in the properties of these populations of jet
knots.

\subsection{Summary of Knot Properties}
\label{sec:summary}

We have measured the radio and X-ray flux density variability in the
knots, the polarization variability, the proper motions and spectral
properties of the 19 radio and 40 X-ray knots.  Here we summarize the
properties of the 13 radio knots with X-ray counterparts, the 6
radio-only knots and the 27 X-ray-only knots.  The following results
and groupings are summarized in Table~\ref{tab:summaryjlg}.

\subsubsection{Knots with Counterparts}

Considering the 13 matched knots, we find that only A3B has a well
determined proper motion in the radio at an apparent velocity of $v/c
= 0.80^{+0.15}_{-0.02}$.  Up until the radio knot A4, the knots with
counterparts are likely stationary with low limits and median
velocities; however, beyond A4 and in the counterjet, the velocities
are inconclusive.  These radio knots are generally less compact and
fainter and so are more affected by artifacts.

Two knot pairs (A1C/AX1C and B1A/BX2) vary in both radio flux density
at 8.4 and 4.8\,GHz and X-ray flux density; of these only A1C varies
in polarization angle.  The increase in radio flux density in A1C is
consistent with what was seen by \citet{hardcastle03jlg} and similar
to the change in X-ray flux density.  

None of these matched knots varies only in the X-ray, while two vary
only in the 8.4\,GHz radio (B2 and S2A); B2 also varies in the degree
of polarization.  A1A and A3A vary in the 8.4\,GHz radio data, but do
not pass our requirement for significant variability in our 4.8\,GHz
data, while S1 varies at both 8.4\,GHz and 4.8\,GHz.  Due to the long
intervals in the 4.8\,GHz data, we cannot rule out that it is
consistent with the 8.4\,GHz radio so only use the 4.8\,GHz data to
determine the spectral indices of the radio knots.

The remaining six pairs of knots vary in neither X-ray nor radio
with two showing a change in the polarization; A2A varies in the angle
of polarization and A1A varies in angle and degree of polarization.
Only two of these knot pairs are completely stable; A5A and SJ3.

\subsubsection{Radio-only Knots}

There are six radio-only knots but the motions of these are mostly
inconclusive with only A1B and A1E having well established velocities
($v/c = 0.53^{+0.06}_{-0.02}$ and $v/c = 0.34^{+0.22}_{-0.15}$
respectively).  The high limits on the remaining velocities cannot
rule out that all of the radio-only knots are moving.  Three
radio-only knots are varying in 8.4\,GHz radio flux density, A1B, A1D,
and SJ1; SJ1 also varies in 4.8\,GHz radio flux density as does A1E.
With regard to the polarization, any changes seen in the moving knots
are attributed to the motion so we detect changes in the angle of
polarization of two knots (A1D and A4) and SJ1 varies in the degree of
polarization.

As mentioned in Section~\ref{sec:counters}, the broad-band spectra of
these knots may be steeper than those of the knots with counterparts,
suggesting a difference in the particle acceleration between these
knots and those with X-ray emission.  However, if we consider that
they may all be moving, this leaves a group of moving, radio-only
knots with half showing signs of radio variability and changes in the
polarization.

The radio-only knots with well established velocities, A1B and A1E,
have some degree of radio flux density variability, while the other
moving knot, A3B, has a steady radio flux density and has already been
described as a group of three sub-regions with only a diffuse X-ray
counterpart.  This X-ray emission may not be associated with the
moving radio material.

\subsubsection{X-ray-only knots}

We detect 27 X-ray knots without radio counterparts and find that the
X-ray light curves for 5 of them show some degree of variability in
their X-ray flux densities, including AX2A and SX1, which may be LMXBs
(Section~\ref{sec:ps}).  The vast majority of these knots therefore
appear stable and many may have flatter X-ray/radio flux density
ratios than those of the knots with counterparts.

We have fitted spectral indices for 14 of these X-ray-only knots and
they are all consistent with synchrotron emission lying in the range
0.58 -- 1.40.  They are also consistent with those measured for the
X-ray knots with radio counterparts (Table~\ref{tab:featuresjlg}).

\begin{table*}
  \centering
    \caption{Summary of X-ray and Radio Knot Behavior}
    \begin{tabular}{lr|ccc|cc|clllll}
      \hline
      \multicolumn{2}{c|}{Name} & \multicolumn{2}{c}{Radio Varying?} & X\-ray & \multicolumn{2}{c|}{Polarization}& \multicolumn{5}{c}{Proper Motion} \\
      X-ray & Radio & 4.8\,GHz & 8.4\,GHz & Varying? & Degree of? & Angle of? & Y/N & median $v$ & upper limit & $\theta_{v}$ & $v_{RA}$ & $v_{dec}$ \\
      & & & & & & & & (c) & (c) & (deg) & (c) & (c) \\
      \hline
      AX1A & A1A  & Y & N & N & Y & Y & N & $0.002$ & $<0.05$ & & $<0.04$ & $<0.00$\\
      ...  & A1B  & N & Y & - & N & N & Y & $0.534^{+0.06}_{-0.02}$ & & $67.40^{+3.70}_{-2.13}$ & $0.50^{+0.06}_{-0.02}$ & $0.21^{+0.01}_{-0.02}$ \\
      AX1C & A1C  & Y & Y & Y & N & Y & N & $0.076$ & $<0.14$ & & $<0.14$ & $<0.11$\\
      ...  & A1D  & N & Y & - & N & Y & I & $0.095$ & $<0.58$ & & $<0.49$ & $<0.30$\\
      ...  & A1E  & Y & N & - & N & N & Y & $0.338^{+0.22}_{-0.15}$ & & $54.18^{+30.12}_{-25.12}$& $0.28^{+0.19}_{-0.08}$ & $0.25^{+0.16}_{-0.16}$ \\
      AX2  & A2A  & N & N & N & N & Y & N & $0.164$ & $<0.24$ & & $<0.09$ & $<0.24$\\
      AX3  & A3A  & Y & N & N & N & N & N & $0.031$ & $<0.05$ & & $<0.03$ & $<0.01$\\
      AX4  & A3B  & N & N & N & N & N & Y & $0.802^{+0.15}_{-0.09}$ & & $71.17^{+11.94}_{-11.76}$& $0.79^{+0.19}_{-0.09}$ & $0.27^{+0.14}_{-0.15}$ \\
      ...  & A4   & N & N & - & N & Y & I & $0.928$ & $<1.00$ & & $<1.00$ & $<0.32$\\
      AX5  & A5A  & N & N & N & N & N & I & $0.016$ & $<0.34$ & & $<0.26$ & $<0.23$\\
      AX6  & A6A  & N & N & N & N & N & I & $0.370$ & $<0.62$ & & $<0.51$ & $<0.36$\\
      BX2  & B1A  & Y & Y & Y & N & N & I & $0.049$ & $<0.74$ & & $<0.69$ & $<0.27$\\
      BX4  & B2   & N & Y & N & Y & N & I & $1.000$ & $<1.00$ & & $<1.00$ & $<1.00$\\
      ...  & SJ1  & Y & Y & - & Y & N & I & $0.371$ & $<0.81$ & & $<0.38$ & $<0.72$\\
      ...  & SJ2  & N & N & - & N & N & I & $0.193$ & $<0.59$ & & $<0.22$ & $<0.55$\\
      SJX1B& SJ3  & N & N & N & N & N & I & $0.213$ & $<0.90$ & & $<0.47$ & $<0.88$\\
      SX1A & S1   & Y & Y & N & Y & N & I & $0.238$ & $<0.83$ & & $<0.79$ & $<0.24$\\
      SX2A & S2A  & N & Y & N & N & N & I & $0.250$ & $<0.37$ & & $<0.30$ & $<0.24$\\
      SX2B & S2B  & N & N & N & Y & N & I & $0.305$ & $<0.60$ & & $<0.55$ & $<0.49$\\[3pt]
      \hline
      \multicolumn{13}{l}{AX2A, EX1, FX1A, GX5 and SX1 vary in the X-ray and do not have radio counterparts.} \\[3pt]
      \multicolumn{13}{l}{AX4A, BX1, BX3, BX5, CX1, CX2, CX3, CX4, EX2, FX1, FX2, FX3, FX5, FX6, FX6A, FX7, GX1, GX2, GX3 and GX4 do not vary in the X-ray and}  \\
      \multicolumn{13}{l}{do not have radio counterparts.} \\ [3pt]
      \multicolumn{13}{l}{The knot proper motions are classified as moving (Y), stationary (N), and inconclusive (I).} \\[3pt]
      \hline
    \end{tabular}
  \label{tab:summaryjlg}
\end{table*}

\begin{table*}
  \centering
    \caption{Normalized radio knot flux densities at 8.4\,GHz with local, on-source background subtraction.}
    \begin{tabular}{lrrrrrrr}
      \hline
           & \multicolumn{7}{c}{Flux (mJy)} \\
      Knot & 1991 & 2002 & 2003 & 2004 & 2006 & 2007 & 2008\\
      \hline
      A1A & $20.72 \pm 0.59$ & $19.97 \pm 1.09$ & $20.98 \pm 1.17$ & $20.98 \pm 1.32$ & $21.76 \pm 1.61$ & $22.94 \pm 1.50$ & $20.67 \pm 1.68$ \\
      A1B & $52.80 \pm 0.77$ & $46.05 \pm 2.24$ & $45.53 \pm 2.26$ & $46.04 \pm 2.61$ & $45.98 \pm 3.09$ & $46.33 \pm 2.76$ & $41.62 \pm 3.18$ \\
      A1C & $34.44 \pm 0.78$ & $38.06 \pm 1.91$ & $38.77 \pm 1.98$ & $39.87 \pm 2.31$ & $41.28 \pm 2.83$ & $40.43 \pm 2.47$ & $41.25 \pm 3.16$ \\
      A1D & $11.18 \pm 0.55$ & $ 8.98 \pm 0.70$ & $ 8.83 \pm 0.75$ & $ 8.23 \pm 0.81$ & $ 8.77 \pm 0.97$ & $ 8.47 \pm 0.91$ & $ 8.46 \pm 0.92$ \\
      A1E & $ 8.25 \pm 0.71$ & $ 9.79 \pm 0.86$ & $10.06 \pm 9.31$ & $ 9.86 \pm 1.03$ & $10.08 \pm 1.20$ & $ 9.09 \pm 1.12$ & $10.27 \pm 1.16$ \\
      A2A & $23.99 \pm 0.80$ & $24.03 \pm 1.37$ & $23.99 \pm 1.43$ & $24.19 \pm 1.62$ & $23.33 \pm 1.86$ & $22.88 \pm 1.69$ & $23.81 \pm 2.00$ \\
      A3A & $ 3.96 \pm 1.29$ & $ 3.75 \pm 1.36$ & $ 4.12 \pm 1.51$ & $ 3.06 \pm 1.54$ & $ 2.88 \pm 1.98$ & $ 4.05 \pm 1.91$ & $ 3.26 \pm 1.56$ \\
      A3B & $22.33 \pm 1.63$ & $21.02 \pm 1.96$ & $20.64 \pm 2.10$ & $18.59 \pm 2.18$ & $16.92 \pm 2.70$ & $18.29 \pm 2.59$ & $18.46 \pm 2.37$ \\
      A4  & $24.75 \pm 1.92$ & $28.35 \pm 2.39$ & $27.48 \pm 2.56$ & $27.17 \pm 2.70$ & $26.44 \pm 3.37$ & $25.98 \pm 3.16$ & $28.72 \pm 3.11$ \\
      A5A & $ 1.08 \pm 0.30$ & $ 1.86 \pm 0.35$ & $ 2.06 \pm 0.34$ & $ 1.23 \pm 0.35$ & $ 1.15 \pm 0.42$ & $ 0.22 \pm 0.51$ & $ 1.81 \pm 0.43$ \\
      A6A & $ 2.85 \pm 0.49$ & $ 2.72 \pm 0.56$ & $ 3.12 \pm 0.54$ & $ 2.04 \pm 0.57$ & $ 1.19 \pm 0.67$ & $ 1.66 \pm 0.83$ & $ 2.41 \pm 0.68$ \\
      B1A & $ 1.59 \pm 0.49$ & $ 2.43 \pm 0.25$ & $ 2.95 \pm 0.55$ & $ 1.25 \pm 0.22$ & $ 1.61 \pm 0.43$ & $ 2.04 \pm 0.12$ & $ 2.15 \pm 0.34$ \\
      B2  & $ 3.86 \pm 0.59$ & $ 4.29 \pm 0.34$ & $ 4.99 \pm 0.67$ & $ 2.46 \pm 0.28$ & $ 5.68 \pm 0.61$ & $ 2.74 \pm 1.38$ & $ 4.05 \pm 0.47$ \\
      SJ1 & $ 5.37 \pm 0.89$ & $12.78 \pm 0.74$ & $10.85 \pm 1.36$ & $14.76 \pm 1.25$ & $16.49 \pm 2.71$ & $18.72 \pm 1.41$ & $20.04 \pm 1.51$ \\
      SJ2 & $ 2.44 \pm 0.74$ & $ 5.13 \pm 0.45$ & $ 4.58 \pm 0.11$ & $ 6.04 \pm 0.88$ & $ 1.71 \pm 0.21$ & $ 3.59 \pm 0.84$ & $ 4.89 \pm 0.48$ \\
      SJ3 & $ 0.55 \pm 0.84$ & $ 1.17 \pm 0.44$ & $ 1.73 \pm 1.20$ & $ 1.53 \pm 2.37$ & $ 1.42 \pm 0.93$ & $ 0.73 \pm 0.37$ & $ 0.77 \pm 0.39$ \\
      S1  & $ 4.28 \pm 1.09$ & $10.10 \pm 0.61$ & $10.26 \pm 1.16$ & $ 6.11 \pm 1.02$ & $ 9.74 \pm 2.01$ & $ 8.95 \pm 1.86$ & $10.40 \pm 0.93$ \\
      S2A & $ 2.28 \pm 0.34$ & $ 1.18 \pm 0.19$ & $ 1.51 \pm 0.39$ & $ 0.16 \pm 0.17$ & $ 1.83 \pm 0.35$ & $ 1.17 \pm 0.81$ & $ 1.94 \pm 0.26$ \\
      S2B & $ 0.82 \pm 0.34$ & $ 0.55 \pm 0.18$ & $ 0.30 \pm 0.38$ & $ 0.05 \pm 0.17$ & $ 1.09 \pm 0.34$ & $ 0.43 \pm 0.81$ & $ 0.61 \pm 0.22$ \\
    \hline
    \end{tabular}
  \label{tab:radiofluxjlg}
\end{table*}

\begin{table*}
  \centering
    \caption{Spectral properties of the Radio Knots}
    \begin{tabular}{lrrrrrrrclll}
      \hline
      \multicolumn{2}{c}{Name} & \multicolumn{2}{c}{Flux Density} & $S_{X}/S_{R}$ & \multicolumn{2}{r}{Radio Spectral Index} & $\alpha_{4.8}^{X}$ & $\alpha_{X}$ & N$_{H}$\\ 
      X-ray & Radio & 8.4\,GHz (mJy) & 1\,keV (nJy) & ($\times10^{-6}$)  & $\alpha_{4.8}^{8.4}$ &  $\alpha_{4.8}^{22.5}$ &  &  & ($\times 10^{22}$\,cm$^{-2}$) \\
      \hline
      AX1A & A1A  & $22.89 \pm 1.21$ & $10.65 \pm 1.64$ & $0.46 \pm 0.07$ & $0.89 \pm 0.34$ & $0.85 \pm 0.21$ & $0.85 \pm 0.14$ & $1.08^{+0.04}_{-0.04}$ & $0.51 \pm 0.02$ \\
      ...  & A1B  & $46.24 \pm 1.08$ & $<19.38$         &                 & $0.84 \pm 0.18$ & $0.80 \pm 0.07$ &                 &  &\\
      AX1C & A1C  & $40.34 \pm 2.27$ & $21.43 \pm 2.97$ & $0.48 \pm 0.09$ & $0.89 \pm 0.22$ & $0.79 \pm 0.14$ & $0.85 \pm 0.23$ & $1.06^{+0.04}_{-0.04}$ & $0.52 \pm 0.02$ \\
      ...  & A1D  & $ 8.45 \pm 0.77$ & $<3.20$          &                 & $1.56 \pm 1.01$ & $0.89 \pm 0.41$ &                 &  &\\
      ...  & A1E  & $ 9.07 \pm 1.00$ & $<1.41$          &                 & $1.66 \pm 1.11$ & $1.42 \pm 0.88$ &                 &  &\\
      AX2  & A2A  & $25.54 \pm 1.35$ & $ 9.29 \pm 0.54$ & $0.39 \pm 0.04$ & $1.57 \pm 0.40$ & $1.16 \pm 0.23$ & $0.86 \pm 0.08$ & $0.77^{+0.07}_{-0.06}$ & $0.55 \pm 0.04$ \\
      AX3  & A3A  & $ 3.12 \pm 0.28$ & $ 4.02 \pm 0.39$ & $1.00 \pm 0.49$ & --              & --              & $0.80 \pm 0.14$ & $0.78^{+0.16}_{-0.15}$ & $0.40 \pm 0.07$ \\ 
      AX4  & A3B  & $19.22 \pm 1.05$ & $ 4.98 \pm 0.28$ & $0.26 \pm 0.04$ & $1.93 \pm 0.59$ & --              & $0.88 \pm 0.08$ & $0.94^{+0.16}_{-0.15}$ & $0.45 \pm 0.06$ \\
      ...  & A4   & $25.92 \pm 2.81$ & $<9.48$          &                 & $1.96 \pm 0.49$ & --              &                 &  &\\
      AX5  & A5A  & $ 1.64 \pm 0.21$ & $ 6.60 \pm 0.39$ & $3.02 \pm 7.01$ & --              & --              & $0.61 \pm 0.10$ & $0.59^{+0.11}_{-0.14}$ & $0.59 \pm 0.07$ \\
      AX6  & A6A  & $ 2.41 \pm 0.34$ & $ 9.39 \pm 0.35$ & $5.64 \pm 2.82$ & $3.91 \pm 2.84$ & --              & $0.70 \pm 0.11$ & $0.51^{+0.04}_{-0.06}$ & $0.51 \pm 0.05$ \\ 
      BX2  & B1A  & $ 2.23 \pm 0.35$ & $19.39 \pm 0.97$ & $9.44 \pm 5.35$ & $1.89 \pm 1.29$ & --              & $0.67 \pm 0.11$ & $0.63^{+0.02}_{-0.02}$ & $0.11 \pm 0.01$ \\
      BX4  & B2   & $ 4.59 \pm 0.46$ & $ 4.94 \pm 0.30$ & $1.80 \pm 0.92$ & $4.88 \pm 2.82$ & --              & $0.77 \pm 0.11$ & $0.91^{+0.06}_{-0.02}$ & $0.08$ \\
      ...  & SJ1  & $18.69 \pm 0.97$ & $<46.79$         &                 &$-0.54 \pm 0.06$ & --              &                 &  \\ 
      ...  & SJ2  & $ 3.58 \pm 0.84$ & $<19.24$         &                 & $1.86 \pm 1.60$ & --              &                 &  \\  
      SJX1B& SJ3  & $ 1.35 \pm 0.26$ & $ 0.99 \pm 0.12$ & $0.65 \pm 0.44$ & --              & --              & $0.83 \pm 0.19$ & $0.50^{}_{}$          & $0.08$\\ 
      SX1A & S1   & $ 9.37 \pm 0.78$ & $ 0.65 \pm 0.06$ & $0.07 \pm 0.02$ & $2.47 \pm 0.84$ & --              & $0.96 \pm 0.12$ & $0.48^{+0.12}_{-0.18}$ & $0.08$ \\
      SX2A & S2A  & $ 1.17 \pm 0.86$ & $ 3.21 \pm 0.21$ & $1.78 \pm 1.26$ & --              & --              & $0.77 \pm 0.58$ & $0.64^{+0.07}_{-0.07}$ & $0.15 \pm 0.04$ \\
      SX2B & S2B  & $ 0.43 \pm 0.85$ & $ 1.88 \pm 0.26$ & $4.14 \pm 7.85$ & --              & --              & $0.75 \pm 2.21$ & $0.77^{+0.20}_{-0.19}$ & $0.13 \pm 0.06$ \\
      \multicolumn{2}{c}{Inner pc-scale jet} & $ 32.13 \pm 4.65  $ & $7.34 \pm 2.41 $ & $0.23 \pm 0.08$ & --  & --  & $0.89 \pm 0.32$ & $0.63^{+0.13}_{-0.11}$ & $0.32 \pm 0.07$ \\
    \hline
    \end{tabular}
  \label{tab:featuresjlg}
\end{table*}

\subsection{Inner Hundred-Parsec-scale Jet}
\label{sec:tiny}

\begin{figure*}
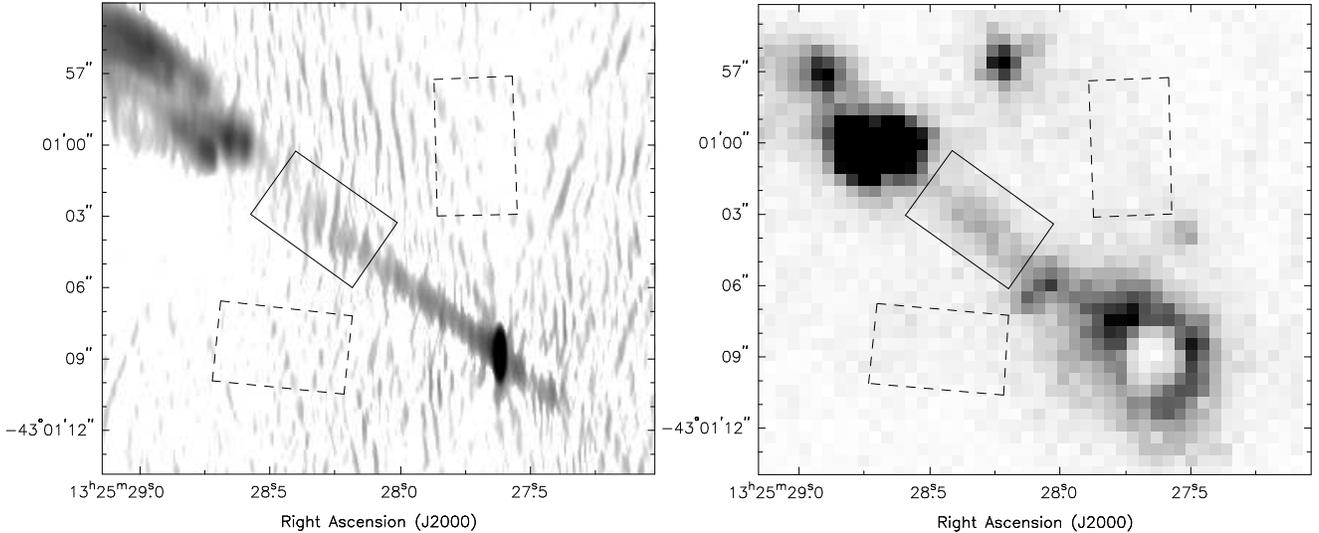

\includegraphics[width=0.48\textwidth]{radio-tinyc2-reg.ps}
\includegraphics[width=0.48\textwidth]{xray-tinyc2-reg.ps}
\caption{Regions used for the analysis of the hundred-parsec-scale inner jet in Cen A (discussed in Section~\ref{sec:tiny}) shown against the 2007 8.4\,GHz radio data (left) and the combined X-ray data in the energy band 0.4 -- 2.5\,keV (right).  The source region is shown in solid line, and the background regions in the dashed line.  These regions were chosen to include as emission from only the hundred-parsec-scale inner jet, excluding the emission from the X-ray point sources near the core.}  
\label{fig:tiny}
\end{figure*}

We detect the inner hundred-parsec-scale jet in both our radio and
X-ray data as a very well collimated feature extending from the core
to the A1 base knots, $\sim250$\,pc downstream of the nucleus.  We
used a rectangular region to isolate the emission from this inner jet
carefully positioned to include as much jet emission as possible
without contamination from the core or the knots; it extends from 6.7
to 12.1\,arcsec (114 -- 204\,pc) from the core as shown in
Figure~\ref{fig:tiny}.  The background emission was estimated from two
regions, positioned such that they extend radially from the nucleus at
the same distances as the jet region, avoiding X-ray point sources.

We measured the X-ray flux densities for each observation and jointly
fitted a power-law with a spectral index of $0.63^{+0.13}_{-0.11}$ and
a Galactic absorption of $0.32^{+0.05}_{-0.02} \times
10^{22}$\,cm$^{-2}$ ($\chi^{2}=134.1$ for 133\,d.o.f.).  This spectral
index is consistent with the spectral indices of the knots downstream;
however, the Galactic absorption is higher in the knots farther
downstream than in this hundred-parsec-scale jet, as expected since
the jet is located within the optical dust lane.  We find that the
X-ray to radio spectral index, $\alpha_{4.8}^{X} = 0.89 \pm 0.08$, is
higher but consistent with that of the base knots while the
X-ray/radio flux density ratio ($0.23 \pm 0.08 \times 10^{-6}$) is
lower than those of the radio knots with X-ray counterparts but higher
than the value for diffuse emission in the center jet \citep[][ see
  Section~\ref{sec:counters}]{hardcastle03jlg}.  We detect no
significant variability in the X-ray flux density; the apparent
fluctuations appear only minor with no obvious trends.  However, the
radio 4.8\,GHz flux density increased by a factor of 2 from 1991 and
shows the first indication of decreasing again in our 2008 data.  We
are unable to measure the radio spectral index for this region of the
jet due to artifacts around the bright core.  Although we detect a
factor of two change in the radio flux density, we cannot make any
firm conclusions on this behavior as this section of the jet is
greatly affected by artifacts from the core.  These results will be
discussed in Section~\ref{sec:dis-tiny}.

\section{Discussion}
\label{sec:dis}

With all of these data we can begin to shed light on the complicated
behavior of the jet: why and where knots are formed, why we can see
them and how they evolve.  In this section we start by examining the
many knot formation models and consider if the behavior of any of the
knots in Cen A supports them.  Not one of these models can explain all
of the observed properties of the knots in Cen A; however, some knots
behave in a way that can be explained by one model or another.  We are
particularly interested to see if the different populations of knots ---
those with counterparts, the radio-only and the X-ray-only knots --- can
all be explained by these models.  We then discuss the effect of
relativistic beaming on the knot emission, which may explain the
observed flux variability, and investigate whether the properties of
the knots can be explained by a spine-sheath model of the jet.  We
also discuss the hundred-parsec-scale jet comparing it to the knots
and to other similar jets.

\subsection{Knot Formation Models}

To determine whether the current models can explain the knots observed
in Cen A, here we compare the observed properties of the knots with
the predicted behavior due to changes in the fluid, namely compression
or rarefaction, or changes in the particle acceleration due to
processes such as reconfinement of the jet, magnetic field reconnection, or
collisions with objects such as molecular clouds and high mass-loss
stars.  

\subsubsection{Adiabatic Compression}
\label{sec:compression}

If a section of the jet's diffuse material underwent adiabatic
compression, the magnetic field, which is frozen into the jet plasma,
would increase in strength as would the number density and energies of
the emitting particles.  This would be reflected in an increase in the
flux density and the break frequency of the synchrotron spectrum, so
we would observe a flatter X-ray to radio spectrum if compression was
responsible for the knots.  \citet{hardcastle03jlg} calculated the
required one-dimensional compression factors, $\cal R$, from the
observed X-ray/radio spectrum using the break frequency ($\nu_{b}
\propto {\cal R}^{4}$ for a tangled field geometry) and considered
whether this level of compression, when applied to the surrounding
diffuse material, is consistent with the observed emission properties.
They found that this level of compression would cause an increase in
the radio volume emissivity of the diffuse material by a factor of
$\sim 10^{11}$ compared to the observed factor of $\sim 2$,
effectively ruling out compression as a creation model for the
X-ray-only knots in Cen A.  Compression in more than one dimension may
reduce the effect to a change in break frequency of $\nu_{b} \propto
{\cal R}^{2}$; however, the change in volume emissivity is still much
higher than the observed factor.  Lesser amounts of compression may
still explain the radio-only knots where the X-ray/radio flux density
ratio of the diffuse material is suppressed, resulting in a X-ray
counterpart too faint to be detected above the diffuse emission, but
they cannot explain any knot with an X-ray counterpart.  In what
follows, we therefore consider only particle acceleration models as
causes for radio knots with X-ray counterparts and X-ray-only knots in
Cen A.  It seems very likely that more than one of these particle
acceleration processes is responsible; here we examine in detail the
predictions of the models and the observed behavior of the knots to
identify those models which are dominant in the jet.

\subsubsection{Impulsive Particle Acceleration}
\label{sec:dis-imp}

If the knots seen in Cen A are the result of impulsive particle
acceleration across the entire knot region, due to a short-lived
processes such as small scale magnetic field reconnection, they would
fade due to synchrotron losses while others would presumably appear in
order to maintain a steady state.  Using the equipartition value of
the magnetic field strength in the A1A knot (Section~\ref{sec:em}), we
would expect a complete change in the appearance of the 1\,keV X-ray
jet emission in $\sim 6$ years; this is not seen.  The X-ray
synchrotron lifetimes of some of the knots that are resolved in the
radio are shown in Table~\ref{tab:synch}.  Consequently, the particle
acceleration processes must be in general long-lived.

The knot HST-1 in M87 may be an impulsive event as it flared
and faded to approximately its original flux in a decade.  The
observed fading is consistent with synchrotron losses; in addition to
a general decrease in all frequencies (X-ray, UV and radio) consistent
with changes in the beaming factor, the X-ray falls-off faster than
the UV or radio \citep{harris09jlg}. However, no knot in Cen A appears
to behave like HST-1; the largest increase in flux is only a factor of
3 over the last 16 years (SJ1), much slower than the flaring of HST-1,
and it has not yet begun to fade.  SJ1 is better described by a
collision model (Section~\ref{sec:dis-moving}).

\subsubsection{Collisions}
\label{sec:dis-col}

A collision between the jet and an obstacle \citep{blandford79jlg}
would result in a local shock complex and is therefore commonly
invoked to explain jet knots.  In this scenario, during the initial
interaction, we would see a steady increase in the luminosity of the
knot relative to the diffuse background.  This is a fast process
relative to the lifetime of the knots, but for plausible obstacle
sizes and speeds is much longer than the period of our observations
and would therefore only be seen as a slight increase in flux.  Once
the obstacle is firmly in the path of the jet, we expect to see a
prolonged period of stable particle acceleration.  Eventually, the
obstacle may be annihilated by the constant impact from the jet fluid;
it could move transversely out of the jet, continuing on its original
path; or it could be carried along the jet, which would cause a
reduction in the shock strength as the obstacle accelerates.  All
these would result in a gradual decrease in the flux and, eventually,
to the complete disappearance of the knot.

In Cen A's jet we can therefore expect to see a range of behaviors for
local shocks in the jet, but the vast majority of knots in this model
are expected to be in a phase of stability with X-ray and radio
emission of a constant flux.  The knots A1A/AX1A and A1C/AX1C are
possibly local shocks due in some part to the reconfinement of the jet
(see Section~\ref{sec:recol}); however, there are many other instances
where there is an X-ray compact source associated with a stationary
compact radio knot: A2A/AX2, A5A/AX5, B1A/BX2, and B2/BX4 in the jet
and SJ3/SJX1b, S2A/SX2A and S2B/SX2B in the counterjet.  There are
some slight changes in the X-ray, radio or polarization in these
systems, but these are not steady, broad-band increases or decreases
which could be attributed to beaming (Section \ref{sec:beam}).  Their
variability may be described as short-term flaring and may be due to
the evolution of the interaction between the jet and the obstacle, to
fluctuations in the jet's fluid flow, or to their shock being curved,
which would be naturally unsteady under small perturbations of the
driving flow.

The majority of the X-ray knots have no detected radio counterparts
and only five of these X-ray-only knots have variability detected in
the X-ray flux density, of which two are probably LMXBs
(Section~\ref{sec:ps}). Of the three probable jet knots, FX1A has a
significant flaring event (X-ray increases by a factor of 2 in the
2002 observation), EX1 shows evidence of a steady decrease in the 2007
VLP observations, which is consistent with predictions of synchrotron
losses, and GX5, a very faint knot, is only detected in 3 of the 10
observations so that we cannot characterize its variability in detail.
Despite these exceptions, the vast majority of these stationary, X-ray
only, compact knots are consistent with a period of stability in the
shock model where the radio counterpart is too faint to be detected in
our data; the range of X-ray/radio ratios we have measured means we
can not dismiss the possibility of faint radio counterparts, and in
fact there is diffuse material emitting in the radio at many of these
positions.

It is still possible that the X-ray-only knots are a separate
population of knots that have flatter spectra than the radio knots with
X-ray counterparts; however, these knots and those with detected
counterparts are consistent with collisions and shock models although
we do not see any knots at a stage where the knots are fading away,
which would be a very short period in the lifetime of the knot
compared to their stable stage.  We do see one knot, AX2A, appearing
in the X-ray during our 2007, $6 \times 100$\,ks observations but we
do not observe a gradual brightening as there is a four-year gap in
the observations prior to the 2007 observations when this may have
occurred, so we cannot say for certain whether this is a new knot or a
LMXB (Section~\ref{sec:ps}).

The radio-only stationary knots may be explained in this scenario by a
weaker shock such as would occur if the obstacle is moving downstream.
As the obstacle is sped up to match the fluid flow, the shock would
weaken until it was too weak to accelerate particles to X-ray emitting
energies.  In this model, the ratio of the numbers of radio and X-ray
knots would be related to their respective lifetimes, but as the
lifetime of the knots also depends on the birth rates, the times taken
for the knots to move along, through or out of the jet, and the
obstacle ablation or acceleration timescales, the relationship would
not be a simple one.  This model explains stationary and very slow
moving knots but those moving at close to the mean jet speed ($v/c
\sim0.5$) cannot be explained by weak shocks, and the limits on the
proper motions of the radio-only knots suggest that many knots may not
be consistent with this model.  A1D is an example of this with an
upper limit speed of $0.58c$. Other models for the radio-only knots
are discussed further in Section~\ref{sec:dis-moving}.

Recent 2.3\,GHz very long baseline interferometry (VLBI) observations
by \citet{tingay09jlg} of the bright A-group knots do not detect our
moving radio knots, A1B and A1E, or A1D while the compact cores of
A1A, A1C and A2A are all resolved. These results strongly support a
collision model for, at least, the stationary radio knots with X-ray
counterparts, and also argue that there is an intrinsic difference
between the stationary and moving knots. If the stationary knots were
due to collisions with an obstacle, we would expect to detect a
compact region where the interaction is occuring; whereas if the
moving knots were due to a non-localized process such as compression
of the fluid flow, we would not expect to detect a compact central
region in the knot.

If we consider the limits on the proper motions of the stationary
knots (A1A; $v/c < 0.05$, A1C; $v/c < 0.14$ and A2A; $v/c <0.24$) it
is reasonable to consider A3A, with a limit of $v/c < 0.05$, to also
be stationary and therefore we can predict that in principle it should
be detectable with VLBI.  The limits on the other knots are all
significantly higher and in two cases, unconfined (A4 and B2) so if
these knots have compact cores that can be detected with VLBI they
could be independently identified as either moving or stationary
knots.  This is particularly interesting in the case of SJ1 as it has
a measured velocity of $v/c = 0.42^{+0.39}_{-0.27}$, but this apparent
velocity due to the effect of artefacts on the observed shape of the
knot.  We cannot resolve a X-ray counterpart due to its proximity to
the core (0.95\,arcsec = 17\,pc), so detecting a compact peak in the
radio with VLBI might give us an independent method of constraining
its motion and therefore the reason for its development.

\citet{tingay09jlg} also detect sub-structure in the knots A1A and
possibly A2A. (The larger-scale structure of these knots is not
detected by the VLBI observations due to the lack of short baselines.)
If we consider A1A, which \citet{tingay09jlg} divide into two compact
sources A1Aa and A1Ab, we find that \citet{tingay09jlg} only detect
approximately half of the flux at 2.3\,GHz (from both of these
substructures) that we would expect based on our observations and
assuming a power-law spectrum.  We must assume that the remaining half
of the emission is coming from more diffuse material on the scale of
0.1--0.2\,arcsec. However, the observations of a compact core reduce
the possible size of the obstacle to an area comparable to the size of
the region detected (0.5 -- 2.5\,pc), which suggests an obstacle such
as O/B stars, which are much more common, rather than more extreme
systems such as Wolf-Rayet stars as suggested by
\citet{hardcastle03jlg}.

{\edit 

As discussed in Section~\ref{sec:res}, we find that many of the X-ray
knots with radio counterparts, and all of the radio-only knots, lie
within the inner arcmin ($\sim 1$\,kpc), while beyond this we find a
complete absence of compact radio knots, even though there is diffuse
radio emission extending to the north inner lobe at $\sim
190$\,arcsec.  It is at $\sim 1$\,arcmin that we also detect a change
in the absorbing column as the jet emerges from the dust lane. Given
the constraints on the geometry of the dust features seen in emission
at $8\mu$m with {\it Spitzer} IRAC \citep{quillen06jlg,quillen08jlg},
it seems unlikely that the jet is interacting with the dust disk
directly beyond the A1 group of knots. However, it remains plausible
that there are a greater number of knots in this inner region due to
collisions with high-mass-loss stars or clumps of cold gas, both of
which will be more common in the central regions of the galaxy.  If we
consider the distribution of stars in Cen A
\citep{vdb76jlg,mellier87jlg} and compare this with the decreasing
number of knots with distance from the core, we find that the number
of stars per unit length in the jet steadily increases with distance
from the nucleus.  This argues that the obstacles are not distributed
like normal stars in the galaxy, but does not rule out the model in
which the obstacles are high mass-loss stars or gas clouds associated
with the central regions of the galaxy.

Farther out, we see a decrease in the spatial density of knots which
is consistent with a predominantly diffuse particle acceleration
mechanism and at $\sim 190$\,arcsec the environment may change again,
as the radio emission expands into a lobe and there is an X-ray
surface brightness discontinuity \citep{kraft08jlg}.  
}

We conclude that describe the majority of knots in the jet of Cen A as
due to the interaction between the jet and an obstacle, including
knots with no detected X-ray or radio counterparts, the exceptions
being those radio-only knots which are moving
(Section~\ref{sec:dis-moving}); however, this assumes that the missing
counterparts exist below the noise level.  If there are really no
counterparts, we require another model to explain the existence of
X-ray-only knots.

\subsubsection{Reconfinement of the Jet}
\label{sec:recol}

It has been suggested \citep[e.g.][]{sanders83jlg} that where the jet
moves from a well-collimated hundred-parsec-scale jet to a complex,
knotty kpc-scale jet, the supersonic fluid encounters a less dense
environment.  It expands into this ambient material and is therefore
likely to cause a reconfinement shock near the boundary of the jet.
This could also be the case if there is a change in the internal
pressure or state of the gas, or because of a change in the external
sound speed or density with no change in the external pressure, which
could occur if the jet is within a relativistic bubble.

At $\sim250$\,pc the inner jet expands from a well-collimated beam to
a diffuse cone of material.  This is indicative of a change in the
ambient pressure which would be consistent with the conditions for a
reconfinement shock.  The base knots A1A/AX1A and A1C/AX1C are
therefore prime candidates for reconfinement shocks.  They are also
stationary in the jet, and A1C is evolving downstream consistent with
this model.  Unfortunately, the presence of the moving knot, A1B,
between two base knots is difficult to explain in a simple
reconfinement model; it would require a unstable knot complex,
possibly ringed, with a shock region that could have been disrupted by
A1B as it moves along the jet.

The knot HST-1 in M87 has also been investigated in terms of a
reconfinement shock model by \citet{stawarz06jlg} and has many traits
similar to A1A and A1C.  It is believed that the stationary, compact,
variable and overpressured flaring region is located immediately
downstream of the point where the reconfinement shock reaches the jet
axis. \citet{stawarz06jlg} also associate the downstream, superluminal
features of HST-1 with a diverging reflected shock.  If we compare
this to Cen A's A1 grouping, then A1A and A1C are consistent with the
flaring region of HST-1 and the fainter downstream components A1D and
A1E, which are moving down the jet, can be compared to the reflection
components.

To summarize, the location of these knots at the point where the jet
widens and the fact that, collectively, they span the width of the jet
are in favor of a reconfinement shock model; however, the fact that
there are two of them and that A1B is apparently moving between them
makes this model harder to accept.

\begin{figure*}
\includegraphics[width=1.0\textwidth]{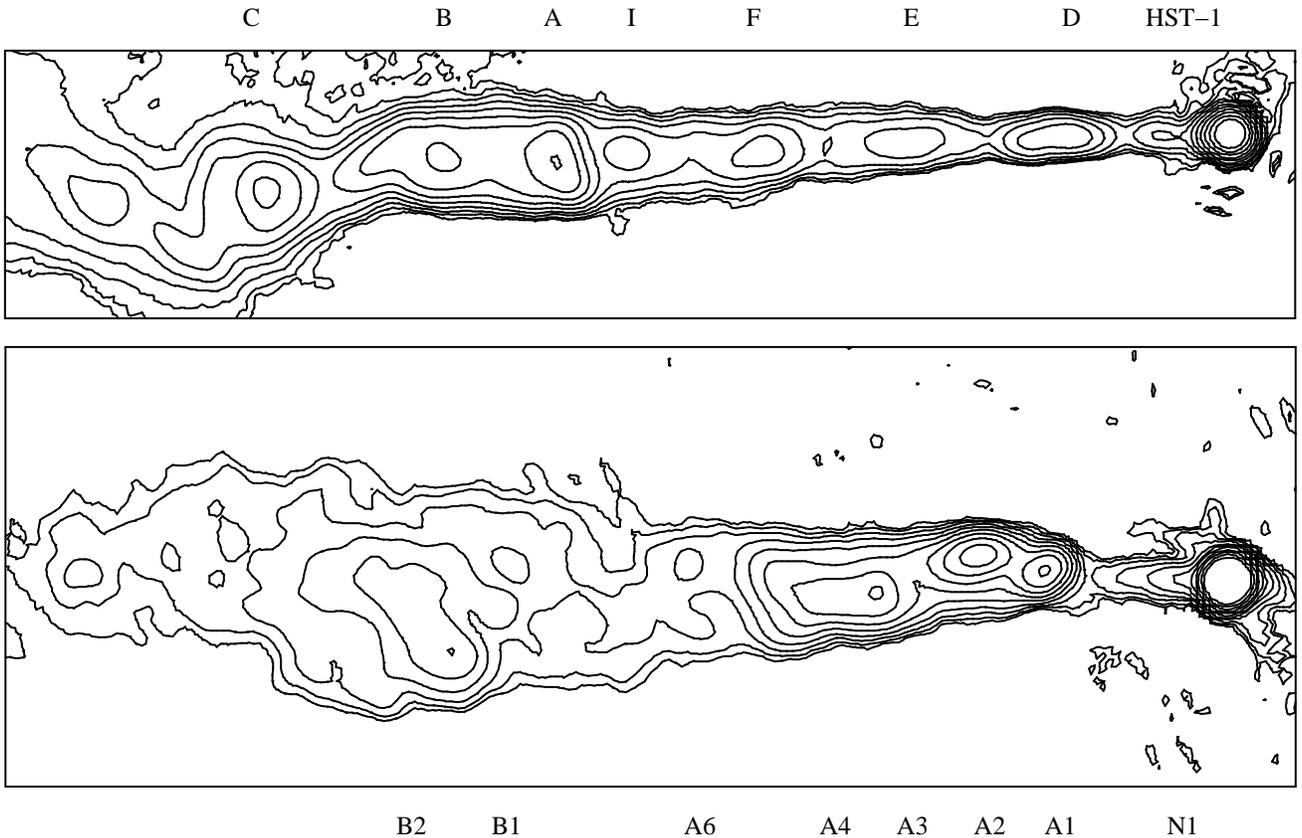}
\caption{M87 (top) and Cen A (bottom) scaled to a common spatial resolution of 33\,pc showing the inner 1.85 kpc of both jets (projected) so that the distances of the knots in each jet can be compared.  The Cen A image is our 2002 A+B 8.4\,GHz radio data and the M87 map is at 5\,GHz \citep{hines89jlg}.  The contours are logarithmic increasing by a factor 2 at each step.}
\label{fig:m87com}
\end{figure*}

\subsubsection{Moving Knots}
\label{sec:dis-moving}

The moving radio knots fit into none of the models discussed
previously (see Section~\ref{sec:dis-col}) as all these models
describe a situation in which the fluid undergoes a change at a
stationary point in the jet.  We also have to consider that there are
no compact X-ray counterparts to the moving radio knots, A1B and A1E,
as well as the diffuse radio emission downstream of A2A; A2B ($v/c =
0.57$), A2C ($v/c = 0.25$) and A2D ($v/c = 0.46$).  The remaining
moving knot, A3B is associated with a clumpy region in the radio jet
rather than a single compact radio knot and has X-ray emission
associated with it, although this could be a projection effect or
emission from a nearby stationary X-ray knot unrelated to the radio
knot.  We must also note that all of the radio-only knots have
  either well-established or inconclusive proper motions so it may be
  that all of the radio-only knots are moving.

Knots that are moving at speeds comparable to the bulk jet flow speed
cannot be due to collisions with a stationary or slow-moving obstacle,
and, as we have argued above (Section~\ref{sec:dis-col}) there is
independent evidence from VLBI observations that there is an intrinsic
difference between the moving and stationary knots in the A group. To
describe the moving knots we require a scenario in which the jet
undergoes a change resulting in a knot of either higher particle
density, higher particle energy, or higher magnetic field, which moves
freely along the jet with the fluid flow. This would be provided by
moderate compression of the jet fluid, as discussed in
Section~\ref{sec:compression}.

Information on X-ray motions of the other knots in the jet will be
invaluable to this problem as the radio-only moving knots may not be
as different as this work suggests, but X-ray proper motions will not
be available for some time.  Our current data spans a sufficient time
frame for motions to be detected; however, the signal-to-noise of the
earlier observations is not good enough to accurately measure the
position of the knots, so that our {\it Chandra} data do not provide useful
constraints.

\subsection{Particle Acceleration Efficiency}
\label{sec:ef}

In models in which the X-ray emission from the knots is synchrotron
emission produced as a result of the interaction between the jet and
an obstacle, the X-ray luminosity places some constraints on the
efficiency of high-energy particle acceleration. Using the constraints
on the knot sizes provided by the VLBI observations
(\ref{sec:dis-col}) we can calculate the fraction of the jet power
intercepted by the knots. Some fraction of this (which cannot exceed
100\%) goes to power the emission at X-ray and other bands.  If we
assume that the knots are in a steady state and that the incident
energy produces electrons with a power-law energy spectrum that
balances the radiative losses, then we can form an inequality: the
ratio of the energy in the X-ray-emitting electrons to that in the
whole electron population must be greater than the ratio of the energy
emitted in the X-rays to the fraction of jet energy intercepted by the
knot.  The unrealistic assumption of 100\% efficient energy transfer
of the absorbed jet energy to relativistic electrons, we find that the
electron energy spectrum has an index, $p \le 2.44$ when we consider
the lowest energy electrons to have $\gamma = 1$.  If we assume that
the energy transfer to the electrons is 10\% efficient, then the
minimum index decreases, $p \le 2.27$ and at 1\% the spectrum is
constrained to have $p \le 2.06$.  However, if the minimum $\gamma$ is
increased to $\gamma=100$, these indices steepen so, for example, 1\%
efficient energy transfer has a minimum index of $p \le 2.13$.  These
calculations use the cross-section and X-ray emission of the knot AX1A
as its size is well constrained \citep{tingay09jlg}.  These minimum
indices put some constraints on the nature of the particle
acceleration in the jet; we note that $p>2.0$, the value expected for
standard acceleration at a non-relativistic strong shock, is possible
unless the efficiency of energy transfer to the electron population
falls below $\sim 1$\%.

\subsection{Beaming}
\label{sec:beam}

If there are changes in the beaming factor in the knots, due to a
change in the velocity or direction of the fluid flow, we would
observe in-step changes in the X-ray and radio flux densities,
assuming all the emission is synchrotron, as beaming to first order is
independent of frequency.  This may occur in stationary or moving
knots as the movement of the fluid through the knot is responsible for
changes in the beaming factor, not the motion of the knot itself.
Although the flaring of HST-1 may be explained as a reconfinement
shock or the result of impulsive particle acceleration, it fades in a
manner best described by beaming combined with synchrotron losses
\citep{harris09jlg}; the emission fades slowly in all frequencies, but
appears to drop-off faster in the X-ray compared to the UV and radio.
Compared to the factor 50 increase in X-ray emission of M87's HST-1,
we observe no flaring events of similar intensity in either our radio
or X-ray data.  The largest increase we observe is in the radio knot
SJ1, which increases in radio emission by a factor of 3.5 over the
last 17 years.  This is the only knot showing variability in the inner
hundred-parsec-scale jet region suggesting it has different physics
from the knots farther down the jet and it is better described by a
collision and shock model.

Smaller increases are seen in other stationary radio knots, increasing
in step in the radio and X-ray although only A1C/AX1C shows a
significant change in both.  In the radio, A1C increases steadily
while in the X-ray we see short-term variability with a dip in the
flux density in the middle of 2007 (Figure~\ref{fig:rlc1}).  This is
consistent with the short-term variability observed in HST-1, which
also show slight dips in the X-ray flux density during an overall
increase.  The knots A1A/AX1A and S1/SX1A also fit these
conditions; however, the subtle increase in flux ($\sim10\%$) is below
the significance limit described in Section~\ref{sec:flux-var}.

We conclude that although we see in-step changes in the radio and
X-ray flux densities in several of the knots, beaming is not the
dominant effect as other mechanisms can explain the observed
behaviors.

\subsection{Spine-Sheath Model}
\label{sec:spine}

Current models of FRI jets \citep{laing02jlg} propose that the jets
have a non-uniform radial velocity profile in which the speed
decreases with increasing radius. Because of early polarization
results that seemed to point to a two-fluid structure, this is often
discussed in the literature in terms of a fast-moving `spine' and
slower `sheath' of the jet. The almost entirely parallel magnetic
field seen in this work and in polarization maps by
\citet{hardcastle03jlg} provides no evidence in itself either for or
against `spine-sheath' models, since modern versions of models with
velocity structure do not predict a transition to a central
perpendicular field in all cases. Some evidence for models with
velocity structure might be provided by the observed localized
edge-brightening in the diffuse material \citep{kataoka06jlg}, if this
is predominantly due to variations in the Doppler enhancement of
different layers of the jet material.  However, this limb-brightening
is not seen along the entire jet; where it is seen, it lies downstream
of a compact knot and could be described as a knot tail.  Similar
structures are seen behind many of the knots, for example A2A, and are
consistent with a shock model with downstream advection
\citet{hardcastle03jlg}.  \citet{kataoka06jlg} also suggest that at
the edges there is a slight hardening of the X-ray emission suggesting
the spectrum changes.

To investigate if this is seen in our deeper X-ray data,
\citet{worrall08jlg} fitted a joint spectrum to all the X-ray knots
that reside in the `inner-spine' and `inner-sheath' regions (at a
distance of 21 to 66\,arcsec from the core with position angles of
$51.2^{\circ}$ -- $57.8^{\circ}$ for the spine and $49^{\circ}$ --
$51.2^{\circ}$ and $57.8^{\circ}$ -- $60^{\circ}$ for the
sheath\footnote{\citet{worrall08jlg} define these pie sections from a
  base position of 13:25:26.98 -43:01:14.06 (not the core)}).  As
described in Section~\ref{sec:spine}, we have spectral fits for 6 of
the 7 X-ray knots in the inner-spine and 3 of the 5 X-ray knots in the
inner-sheath regions defined by \citet{worrall08jlg}.  The weighted
mean of the spectral indices we measure for the X-ray knots in these
regions are $\alpha_{spine} = 0.62 \pm 0.01$ and $\alpha_{sheath} =
0.89 \pm 0.05$ respectively, which are consistent with the results of
Worrall et al.  but not with each other.  On closer inspection of the
individual knot spectral indices, we find that the inner-spine
spectral index is dominated by the bright X-ray knot BX2 ($\alpha_X =
0.67 \pm 0.11$).  Given the small number statistics and the dominance
of individual knots, it is difficult to draw conclusions about the
behavior of either knot population.

As discussed in Section~\ref{sec:pm}, we see no dependence of the
velocities on the angular position of the knots, nor on the distance
of the knots from the core, although our sample is too small for a
statistical analysis.  We have considered the directions of motion and
find that they are consistent with following the fluid flow, appearing
to move toward the downstream regions of bright material.  The
directions of motion are not consistent with moving exactly parallel
to the jet axis but we argue that the ridge line through the jet is
not at a constant positional angle and that the fluid flow is
complicated and not a simple laminar flow directed away from the core.
We therefore cannot comment further on the possibility of a faster
moving spine in the jet of Cen A nor on the possible migration of the
knots towards or away from the jet axis.

\begin{figure}
\includegraphics[width=0.35\textwidth,angle=270]{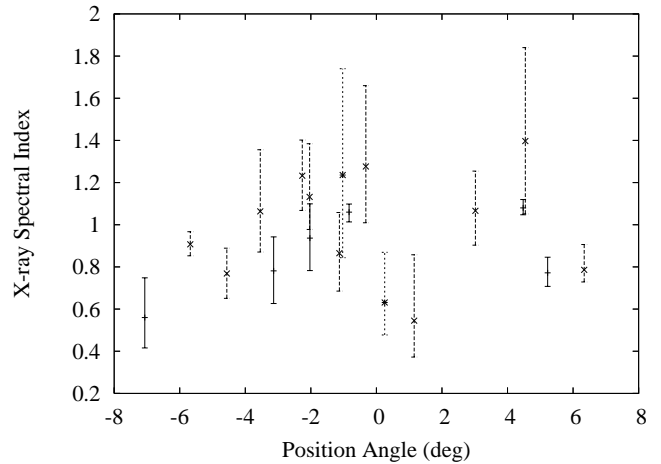}
\caption{The X-ray photon index of knots in the jet of Cen A as a function of angular position measured from the centre of the jet at a position angle of $54.0^{\circ}$, from the core to 60\,arcsec (solid line), from 60 -- 190\,arcsec (dashed line) and farther than 190\,arcsec from the core (dotted line).}
\label{fig:angsections}
\end{figure}

\subsection{Inner Hundred-Parsec-scale Jet}
\label{sec:dis-tiny}

We found that the spectrum of the inner hundred-parsec-scale jet of
Cen A is flatter in the X-ray than the base knots and the diffuse
material \citep{hardcastle03jlg} but consistent in the radio to X-ray
spectral index.  These measurements are also consistent with the
values for knots farther along the jet.  The X-ray/radio flux density
ratio is lower for the inner hundred-parsec-scale jet than for any of
the knots, except S1/SX1A; however, it is higher than the X-ray/radio
flux density ratio of the diffuse material farther down the jet
suggesting that the inner jet is more efficient at making X-rays for a
given amount of radio than the diffuse emission farther up the jet.
This is consistent with what is seen in other FR\,I jets such as that in
3C\,66B \citep{hardcastle01jlg}. 

\section{Summary and Conclusions}
\label{sec:con}

Our results can be summarized as follows:

\begin{itemize}

\item We rule out impulsive particle acceleration in the knots of Cen
  A as we detect no extreme variability in the X-ray knots, in
  contrast to what is seen in knot HST-1 in M87.  We see essentially
  the same distribution of X-ray knots in our most recent observation
  as was seen in the earliest {\it Chandra} observations in 1999.
  This would not be the case if the knots were impulsive as they would
  fade due to synchrotron losses indicating long-lived particle
  acceleration in the knots of Cen A.

\item For those radio knots with X-ray counterparts, the most likely
  formation mechanism is a collision between the jet and an obstacle,
  resulting in a local shock.  We see no significant variability in
  many of these knots, suggesting a long-lived, stable stage of
  particle acceleration during the interaction between the jet and the
  obstacle.

\item The formation of knots at the point where the inner
  hundred-parsec-scale jet broadens abruptly suggests that these base
  knots (A1A and A1C) may be reconfinement shocks; however, this is
  complicated by the presence of a radio-only knot (A1B) moving
  downstream between the possible confinement-shock knots.

\item We detect a factor of 3 increase in radio flux density of the
  counterjet knot SJ1.  This knot lies only 17\,pc from the nucleus so
  is unresolved in the X-ray; however, it was still increasing in flux
  in the most recent observation (Dec 2008) so we plan to continue to
  monitor its radio behavior with the VLA.

\item We detect proper motions in three of our radio knots; two of
  which have no compact X-ray counterparts and a third which has only
  diffuse X-ray emission associated with it.  Studies of the
  distribution of the moving knots are inconclusive due to the low
  number of well-established proper motions; however, that the
  direction of motion of the knots may not be directly parallel to the
  jet axis which appears to varies along the jet.  Their motions are
  all downstream and they show no dependence on the position of the
  knot within the jet.

\end{itemize}

The most likely cause of knots in the jet is collisions; if the
X-ray-only knots have faint radio counterparts and the radio-only
knots are seen only during the latter stages of the collision when the
interaction is weaker, then only the moving knots are a separate
population.  These may include all the radio-only knots but our proper
motion measurements are inconclusive for many of these.  We argue that
these moving knots are due to compressions in the fluid flow that do
not result in particle acceleration to X-ray emitting energies.  It is
possible, however, that the X-ray-only knots are also a separate
population with flatter X-ray to radio spectra than those with
counterparts, in which case we currently have no model for their
formation.

Compared to other FR\,I jets, Cen A is atypical, with an obscuring
dust lane extending out to 1\,kpc from the core which greatly affect
the jet and its knots.  Other galaxies where dust has been detected,
such as 3C31 and 3C449, have much smaller disks, which cannot affect
even the innermost regions of the observed X-ray jet.  If we can
attribute the knot dominated particle acceleration of the inner kpc to
the presence of this disk then we can postulate that the X-ray jet
emission seen in other FR\,I galaxies should be comparable to the
dominant diffuse particle acceleration that dominates farther out in
the Cen A jet.  We would then predict that knot-dominated structure
will not be seen in other FR\,I galaxies.

\vspace{-10pt} \acknowledgements We gratefully acknowledge financial
support for this work from the STFC (research studentship for JLG),
the Royal Society (research fellowship for MJH), and NASA (grant
GO7-8105X to RPK). The National Radio Astronomy Observatory is a
facility of the National Science Foundation operated under cooperative
agreement by Associated Universities, Inc.  We acknowledge helpful
comments from an anonymous referee.

{\it Facilities:} \facility{CXO (ACIS)}, \facility{VLA}

\appendix

\section{Flux Variability Plots}

\begin{figure*}
\includegraphics[width=1.45\textwidth,angle=90]{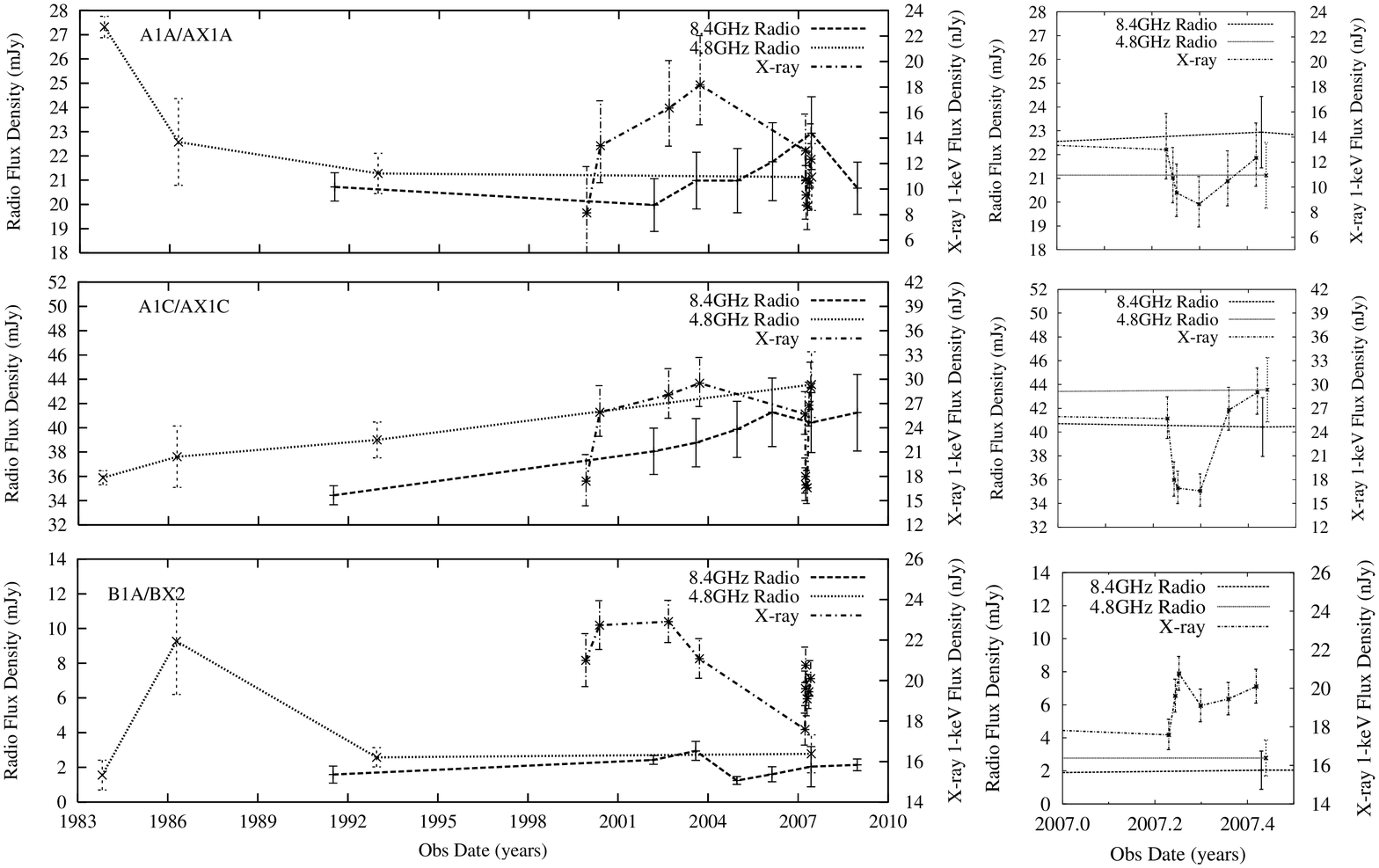}
\caption{X-ray and 8.4\,GHz radio light curves for the radio knots in Cen A, which show some degree of radio or X-ray variability AND have X-ray counterparts; A1C (top panel), B1A, (middle panel) and B2 (bottom panel).  The right hand images are cropped to highlight changes in the X-ray flux density during the {\it Chandra} VLP observations.}
\label{fig:lc1}
\end{figure*}

\begin{figure*}
\includegraphics[width=1.45\textwidth,angle=90]{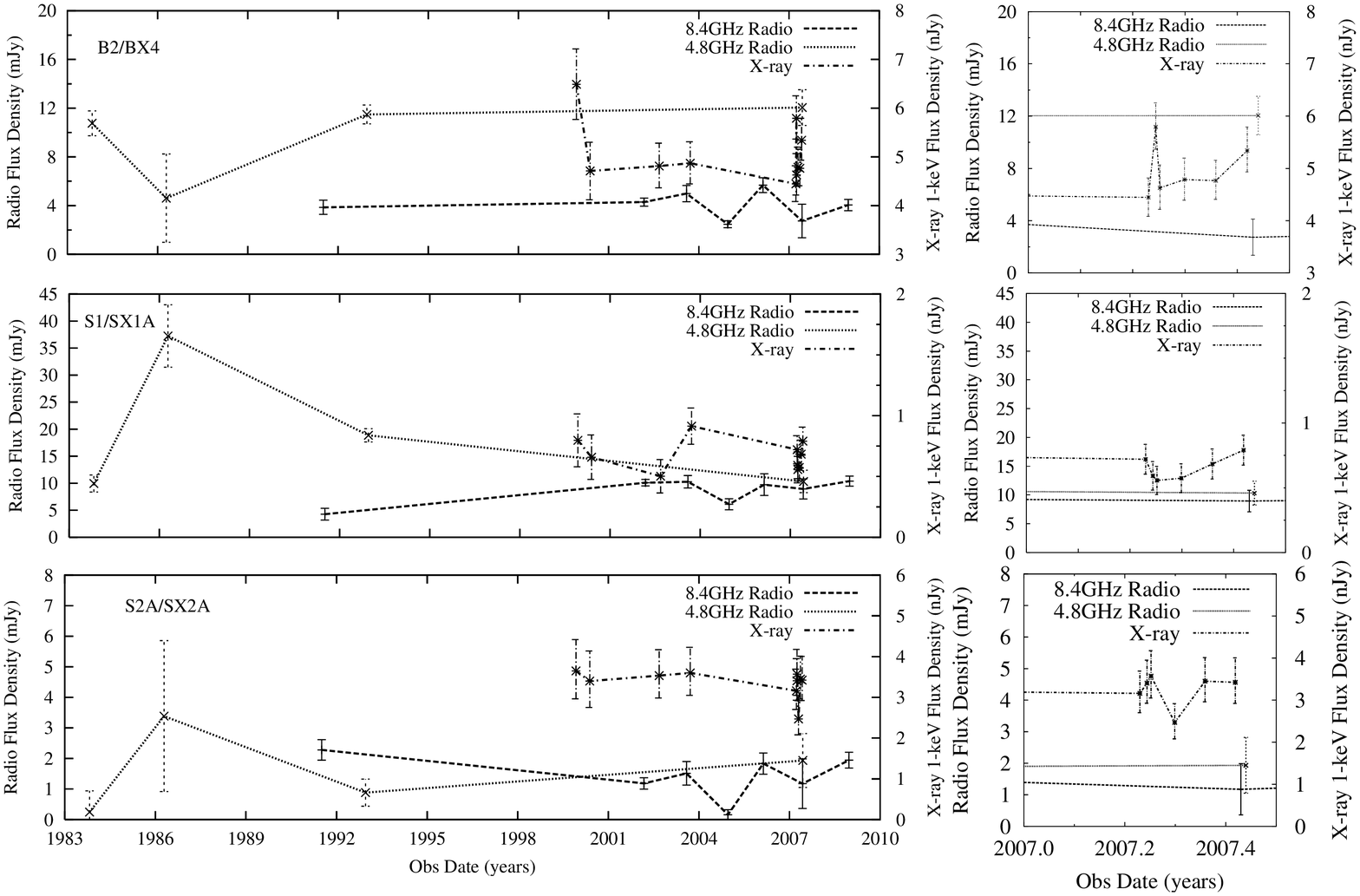}
\caption{X-ray and 8.4\,GHz radio light curves for the radio knots in Cen A, which show some degree of radio or X-ray variability AND have X-ray counterparts; S1 (top panel), S2A (middle panel) and S2B (bottom panel). The right hand images are cropped to highlight changes in the X-ray flux density during the {\it Chandra} VLP observations.}
\label{fig:lc2}
\end{figure*}

\begin{figure*}
\includegraphics[width=1.0\textwidth]{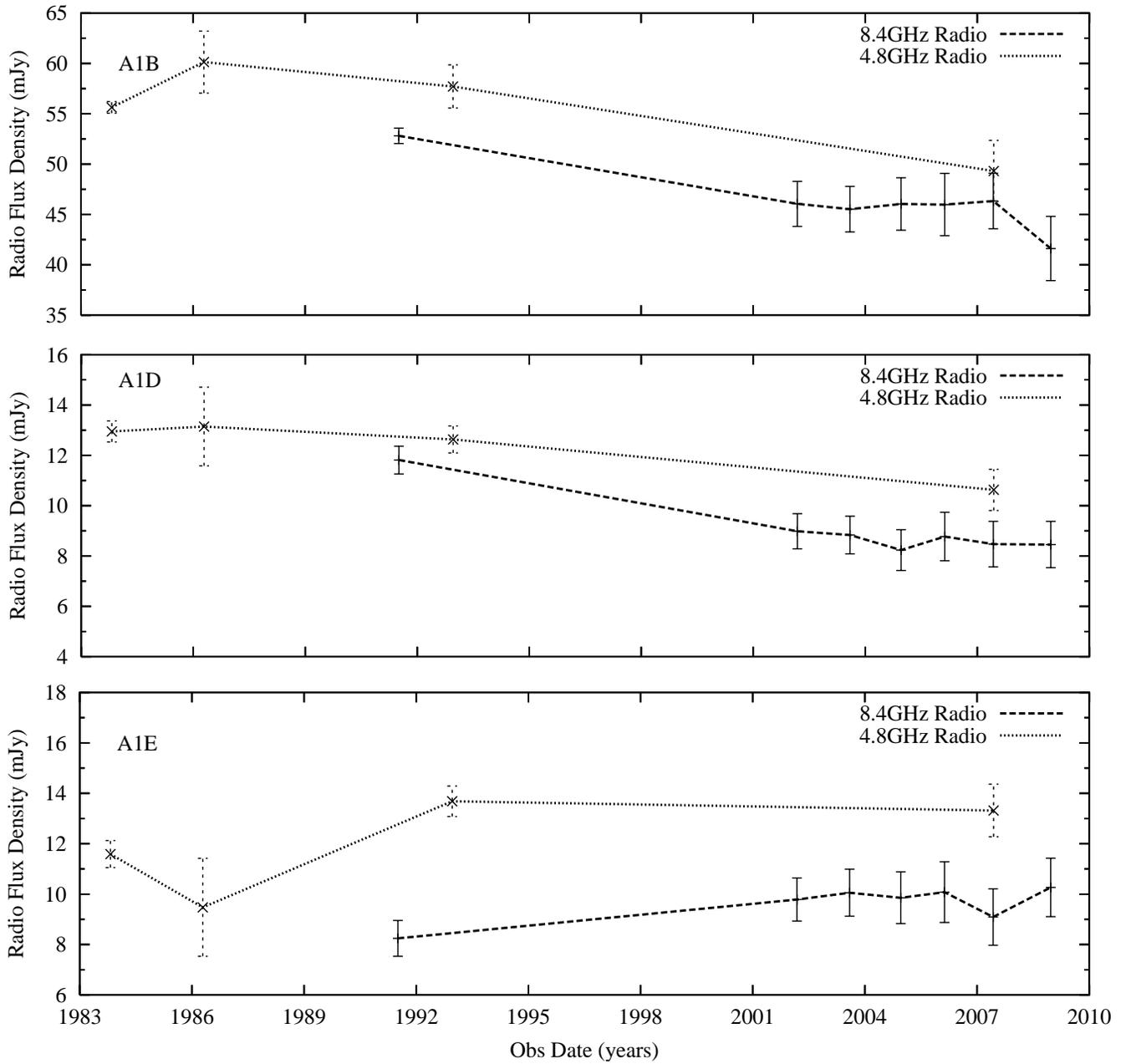}
\caption{8.4\,GHz and 4.8\,GHz radio light curves for the radio-only knots in Cen A; A1B (top panel), A1D (middle panel) and A1E (bottom panel).}
\label{fig:rlc1}
\end{figure*}

\begin{figure*}
\includegraphics[width=1.0\textwidth]{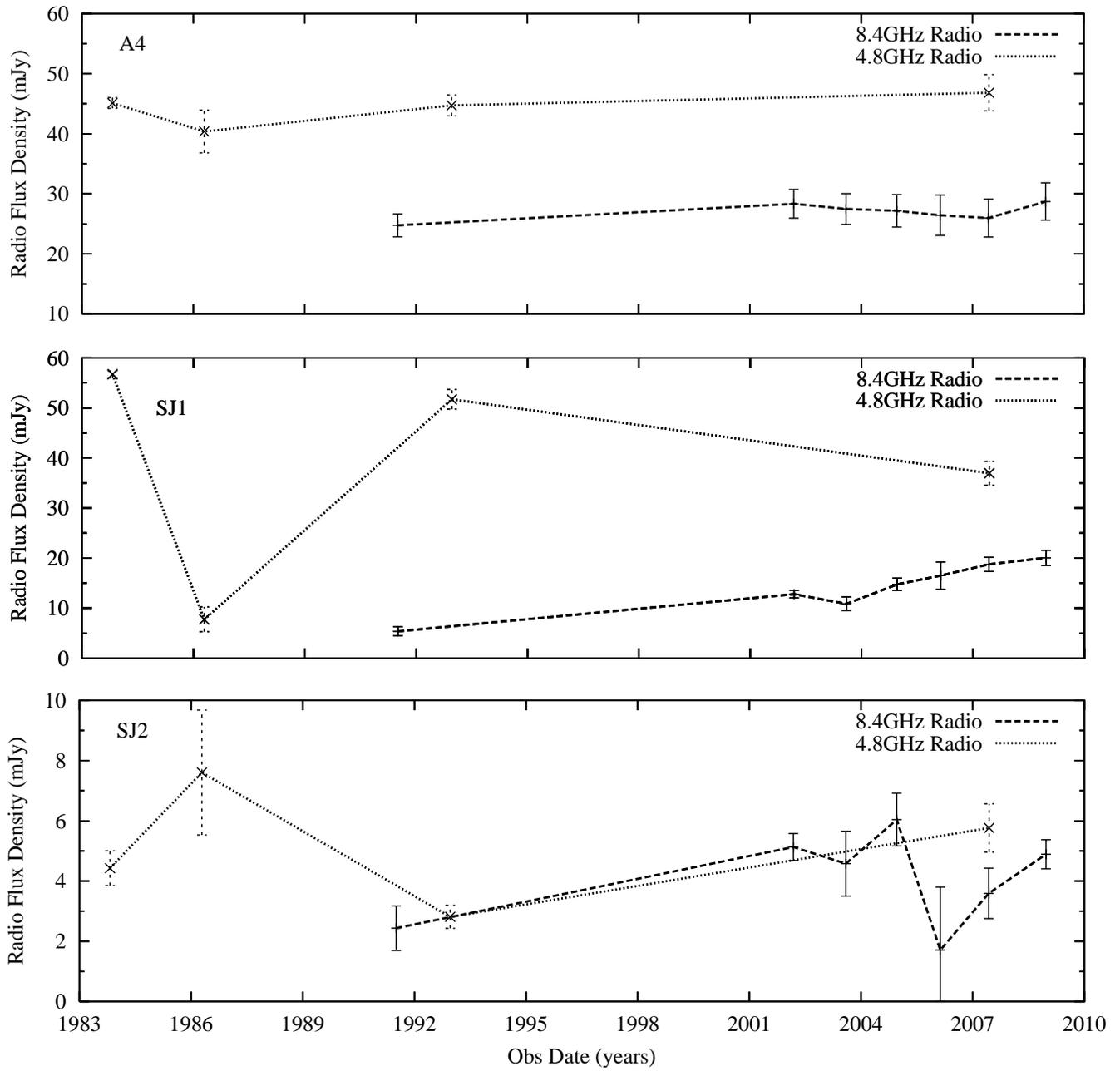}
\caption{8.4\,GHz and 4.8\,GHz radio light curves for the radio-only knots in Cen A; A4 (top panel), SJ1 (middle panel) and SJ2 (bottom panel).}
\label{fig:rlc2}
\end{figure*}

\begin{figure*}
\includegraphics[width=1.0\textwidth]{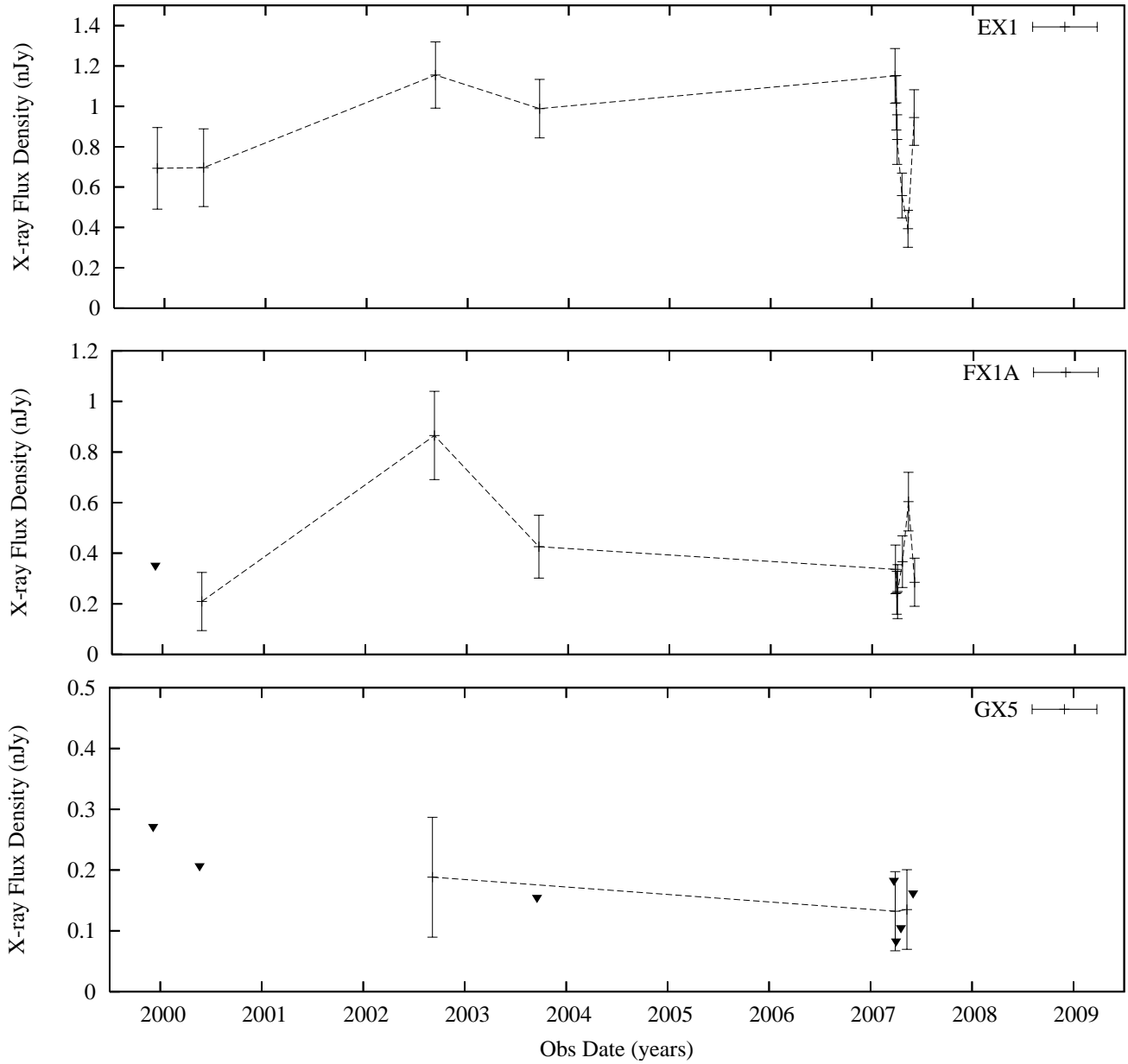}
\caption{X-ray light curves for the X-ray only knots in Cen A, which show signs of variability; AX2A (top left), EX1 (top right), FX1A (bottom left) and GX5 (bottom right).}
\label{fig:xlc}
\end{figure*}

\begin{figure*}
\includegraphics[width=1.0\textwidth]{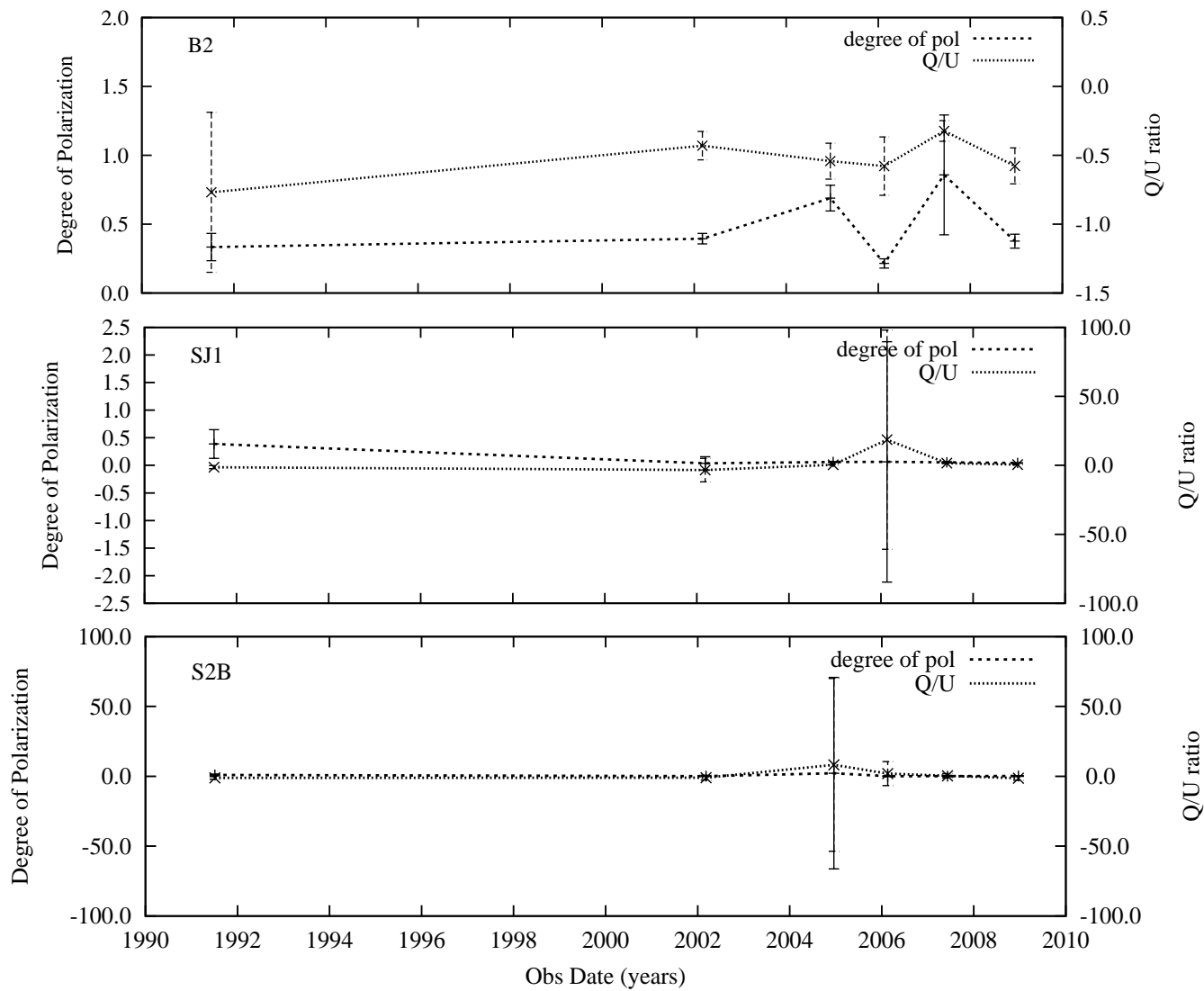}
\caption{Plots for the radio knots in Cen A, which show significant variability in the degree of polarization only with the Degree of Polarization (dotted lines) and the Angle of Polarization (dashed lines): B2 (top panel), SJ1 (middle panel) and S2B (bottom panel).}
\label{fig:poldeg}
\end{figure*}

\begin{figure*}
\includegraphics[width=0.262\textwidth,angle=270]{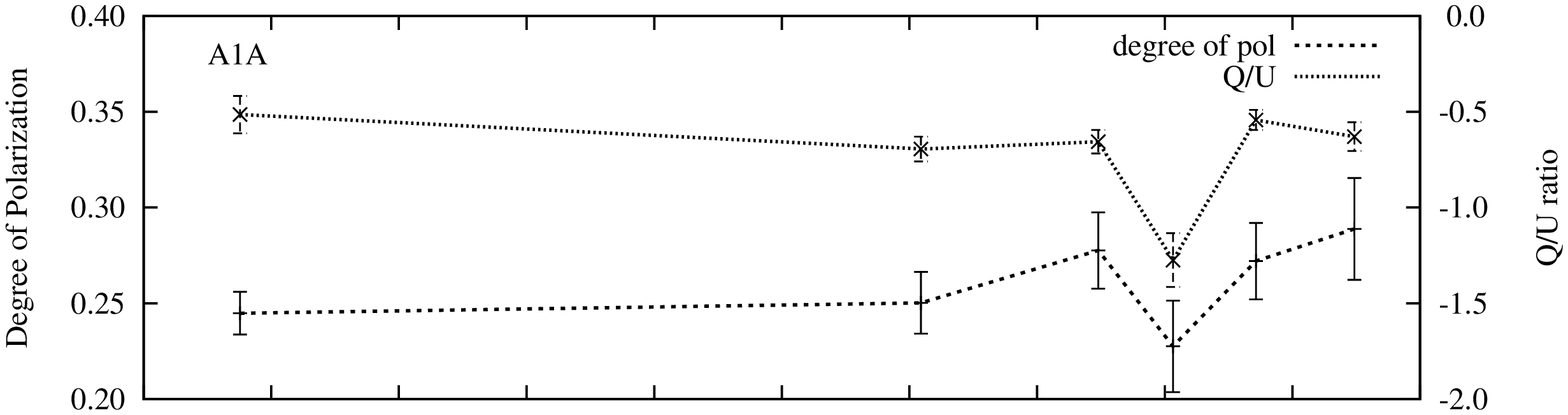}\\
\includegraphics[width=0.262\textwidth,angle=270]{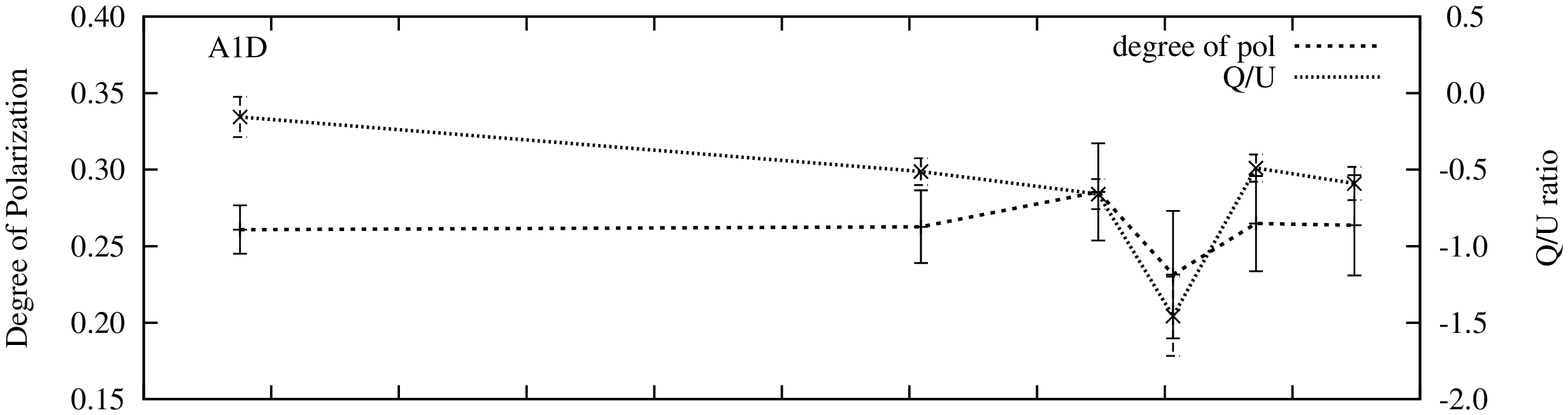}\\
\includegraphics[width=0.317\textwidth,angle=270]{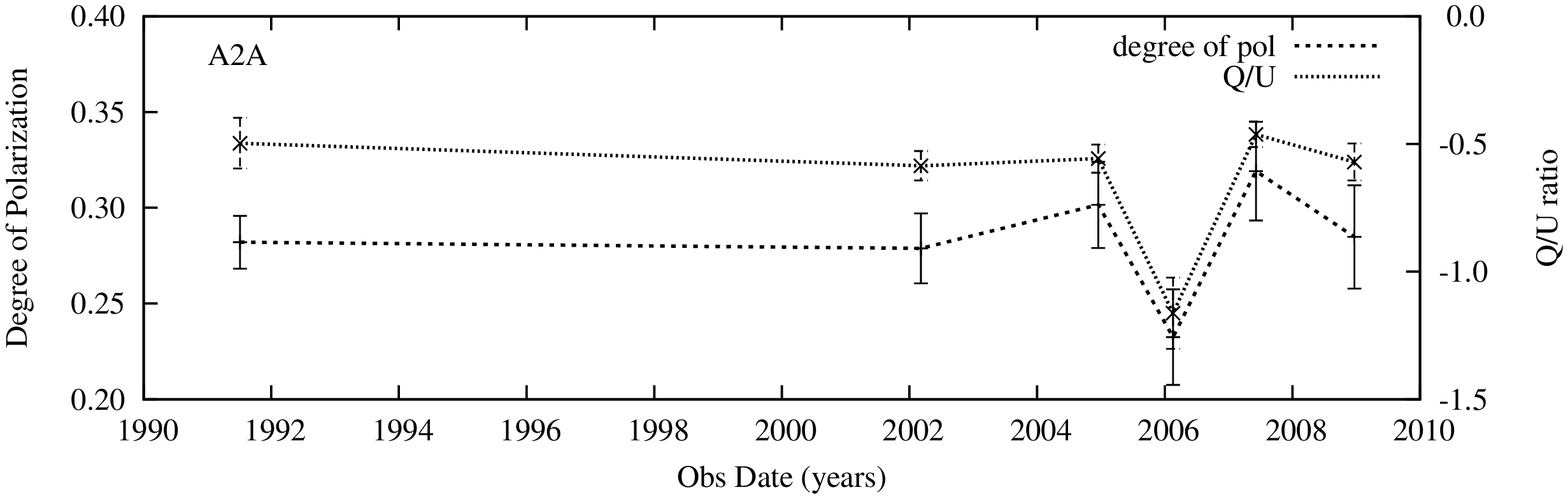}\\
\caption{Plots for the stationary radio knots in Cen A, which show significant variability in the angle of polarization only with the Degree of Polarization (dotted lines) and the Angle of Polarization (dashed lines): A1A (top panel), A1C (middle panel) and  A2A (bottom panel).}
\label{fig:polang}
\end{figure*}

\end{document}